%% file: ms.tex
\documentclass[12pt,preprint]{aastex}

\usepackage{graphicx,natbib,amsmath,amssymb}
\bibpunct[, ]{(}{)}{;}{a}{}{,}
\newcommand{\swift}{\textit{Swift}}
\newcommand{\hst}{{\it HST}}

\slugcomment{ApJ 756, 187 (2012)}

\shorttitle{TOUGH1: Sample and Catalogs}
\shortauthors{Hjorth et al.}

\begin{document}

\title{
The optically unbiased GRB host (TOUGH) survey.
I. Survey design and catalogs%
\thanks{Based on observations collected at the European Southern Observatory, 
Paranal, Chile (ESO Large Programme 177.A-0591). Reduced data and catalogs are 
made available at \texttt{http://dark-cosmology.dk/TOUGH} and
\texttt{http://archive.eso.org}.}
}


\author{ 
Jens~Hjorth\altaffilmark{2}, 
Daniele~Malesani\altaffilmark{2}, 
P\'all~Jakobsson\altaffilmark{3}, 
Andreas~O.~Jaunsen\altaffilmark{4}, 
Johan~P.~U.~Fynbo\altaffilmark{2}, 
Javier~Gorosabel\altaffilmark{5}, 
Thomas~Kr\"uhler\altaffilmark{2}, 
Andrew~J.~Levan\altaffilmark{6},
Micha{\l}~J.~Micha{\l}owski\altaffilmark{7}, 
Bo~Milvang-Jensen\altaffilmark{2}, 
Palle~M\o ller\altaffilmark{8}, 
Steve~Schulze\altaffilmark{3}, 
Nial~R.~Tanvir\altaffilmark{9},
and Darach~Watson\altaffilmark{2}
}


\altaffiltext{2}{Dark Cosmology Centre, Niels Bohr Institute, University of
Copenhagen, Juliane Maries Vej 30, DK-2100 Copenhagen \O, Denmark; 
jens@dark-cosmology.dk} 
\altaffiltext{3}{Centre for Astrophysics and Cosmology, Science Institute, 
University of Iceland, Dunhagi 3, 107 Reykjav\'ik, Iceland}
\altaffiltext{4}{Institute of Theoretical Astrophysics, University of Oslo, 
PO Box 1029 Blindern, NO-0315 Oslo, Norway}
\altaffiltext{5}{Instituto de Astrof\'\i sica de Andaluc\'\i a 
(IAA-CSIC), P.O. Box 03004, E-18080 Granada, Spain}
\altaffiltext{6}{Department of Physics, University of Warwick, 
Coventry CV4 7AL, UK}
\altaffiltext{7}{SUPA, Institute for Astronomy, University of Edinburgh,
Royal Observatory, Edinburgh, EH9 3HJ, UK}
\altaffiltext{8}{European Southern Observatory, Karl-Schwarzschild-Str.\ 2, 
D-85748 Garching by M\"unchen, Germany}
\altaffiltext{9}{Department of Physics and Astronomy, 
University of Leicester, University Road, Leicester LE1 7RH, UK}

\begin{abstract}

Long-duration gamma-ray bursts (GRBs) are powerful tracers of star-forming
galaxies.
We have defined a homogeneous
subsample of 69 \swift\ GRB-selected galaxies spanning a very 
wide redshift range.
Special attention has been devoted to making the sample optically 
unbiased through simple and well-defined selection criteria based on the 
high-energy properties of the bursts and their positions on the sky. 
Thanks to our extensive follow-up observations, this sample has now achieved a 
comparatively high degree of redshift completeness, and thus provides a legacy 
sample, useful for statistical studies of GRBs and their host galaxies. In 
this paper we present the survey design and summarize the results of our 
observing program conducted at the ESO Very Large Telescope (VLT) aimed at 
obtaining the most basic properties of galaxies in this sample, including a 
catalog of $R$ and $K_{s}$ magnitudes and redshifts. 
We detect the host galaxies for
80\% of the GRBs in the sample, although
only 
42\% have
$K_{s}$-band detections, which confirms that GRB-selected host galaxies 
are generally blue.
The 
sample is not uniformly blue, however, with two extremely red objects 
detected. Moreover, galaxies hosting GRBs with no optical/NIR afterglows, whose identification therefore
relies on X-ray localizations, are 
significantly brighter 
and redder 
than those with an optical/NIR afterglow. 
This supports a scenario where 
GRBs occurring in more massive and dusty galaxies frequently suffer high optical obscuration.
Our spectroscopic campaign 
has resulted in 
77\%
now having redshift measurements, with a 
median redshift of $2.14\pm0.18$. TOUGH alone includes 17 detected 
$z>2$
\swift\ GRB host galaxies suitable for individual 
and statistical studies---a substantial increase over previous samples. 
Seven hosts have detections of 
the Ly$\alpha$ emission line and we can exclude an early indication that 
Ly$\alpha$ emission is ubiquitous among GRB hosts, but confirm that Ly$\alpha$ 
is stronger in GRB-selected galaxies than in flux-limited samples of 
Lyman break galaxies.

\end{abstract}





\section{Introduction\label{introduction}}

Following the discovery of long-duration gamma-ray burst (GRB) optical 
afterglows \citep[e.g.,][]{2000ARA&A..38..379V}, evidence rapidly accumulated 
pointing to their origin in
the deaths of massive stars \citep{1998Natur.395..670G,2002AJ....123.1111B,2003ApJ...591L..17S,2003Natur.423..847H}.
This soon led to their being touted as a potentially powerful
means of discovering and reckoning star formation activity, 
which could bypass many of the biases that hamper
other star formation measures
\citep[e.g.,][]{1998MNRAS.294L..13W}. The detection of GRBs is at once both 
independent of the luminosity of their host galaxy and, 
at high energies, unaffected by dust obscuration. They are also extremely 
luminous at all wavelengths, so that they can be used to perform transmission 
spectroscopy of their host galaxies, even where such hosts are too faint to 
be detected photometrically \citep[e.g.,][]{2010A&A...523A..70T}. This also 
permits them to be found across the universe from very low to very high 
redshift: indeed GRB\,090423 holds the record at $z=8.2$ for the most distant
spectroscopically confirmed object known 
\citep{2009Natur.461.1254T,2009Natur.461.1258S}. 

The history of baryonic structure in the universe to a large extent revolves 
around the co-evolution of stars, galaxies, metals, and dust.  In recent years, 
a range of methods has been applied to the identification and study of 
different populations of high-redshift galaxies, but all are variously 
affected, 
to a greater or lesser extent, by the problems of 
dust extinction, flux-limited samples, source confusion, and incomplete 
redshift determinations. 
Notable examples are:
Lyman break galaxies \citep[LBGs;][]{2003ApJ...592..728S}, sub-mm galaxies
\citep[SMGs; e.g.,][]{2000AJ....119.2092B,2002PhR...369..111B}, Ly$\alpha$ 
emitters \citep[LAEs; e.g.,][]{2003A&A...407..147F,2005ASSL..329..293F,2006ApJ...642..653S},
distant red galaxies \citep[DRGs;][]{2004ApJ...617..746D}, 
damped Ly$\alpha$ absorbers 
\citep[DLAs;][]{2002ApJ...574...51M,2005ARA&A..43..861W}, and
high-redshift dropout galaxies \citep[e.g.,][]{2010ApJ...709L.133B}. 
Studies of these populations have led to major advances in our understanding
of galaxy evolution, but our view remains far from complete, particularly at 
higher redshifts.

The major stumbling blocks to the routine use of GRBs as 
tools for the investigation of cosmic history have been the relatively slow 
buildup of statistically useful samples, and the disconnect between the 
high-energy properties of the burst and parameters useful for measuring the 
host and its star formation rate (SFR).  These latter parameters, such as the 
redshift, the host-galaxy bolometric luminosity, metallicity, and dust content
can only be determined from longer wavelength observations in the UV, optical, 
infrared, millimeter, and radio regimes.

Accurate X-ray localizations, as an intermediate step between the gamma-rays
and the optical/infrared, have turned out to be critical to finding the
afterglows and the host galaxies of GRBs. The first
detections provided by 
\emph{BeppoSAX} \citep{1997Natur.387..783C} heralded the afterglow era.  
Since 2005, \swift\ \citep{2004ApJ...611.1005G}, with its autonomous slewing capability 
and onboard X-Ray Telescope \citep[XRT,][]{2005SSRv..120..165B} has been
extremely successful in this regard, delivering high-accuracy (few arcseconds) 
positions for approximately a hundred bursts every year. It has become the GRB 
facility par excellence thanks to its innovative, intelligent design, open 
access policy, energetic team, and active
community support.

In spite of \swift's success, some of the problems that beset most other 
star-formation indicators also affect GRBs. Despite
many attempts to infer distances from the prompt high-energy properties 
\citep[e.g.,][]{2005MNRAS.364..163G},  
UV/optical/NIR observations (and preferably spectroscopy) 
remain the only route to  
reliably find the redshift of a GRB or its 
host, and we must observe the hosts of GRBs directly across a range of 
wavelengths to connect SFR to GRB rate. Thus, to avoid optical 
biases in GRB selection
requires a sample as complete as possible in redshift and host-galaxy
properties.
\swift's excellent X-ray localization completeness and large sample size 
provide a first opportunity for this to be attempted.

Several previous studies of GRB hosts have found that the majority
of GRB hosts are faint, blue galaxies 
\citep{1999ApJ...519L..13F, 2003A&A...400..499L, 2004A&A...425..913C}
with low stellar masses 
\citep{2006ApJ...653L..85C,2010ApJ...721.1919C,2009ApJ...691..182S}.
The GRBs generally occur within the UV-bright parts of their hosts 
\citep{2002AJ....123.1111B,2006Natur.441..463F},
consistent with their association with star formation. 
They are also typically fainter and have lower metallicities than the galaxies 
selected in Lyman-break surveys in the same 
redshift range \citep{2005MNRAS.362..245J,2006NJPh....8..195S,2008ApJ...683..321F}. 
Accounting for systematic incompleteness biases due to dust obscuration,
i.e., the so-called dark burst problem
\citep{2001A&A...373..796F,2009ApJS..185..526F,2003ApJ...592.1018D,2004ApJ...617L..21J,2009AJ....138.1690P,2011A&A...534A.108K,2012MNRAS.421...25S,2012arXiv1202.1434R,2012MNRAS.421.1265M},
has also been attempted with some success
\citep[e.g.,][]{2006A&A...447..897J,2009ApJS..185..526F}. 

Submillimeter (submm)
and radio surveys of GRB hosts have shown that only a few could be 
ultraluminous infrared galaxies 
\citep[][]{2003ApJ...588...99B}, indicating that obscured star formation is 
not dominant in GRB-selected galaxies \citep{2004MNRAS.352.1073T}. Moreover, 
the candidate submm bright GRB hosts are bluer, hotter, and at lower 
redshift than typical of SMGs 
\citep[e.g.,][]{2003ApJ...588...99B,2008ApJ...672..817M}. 

Most of the studies mentioned above have been limited to fairly low-redshift 
samples ($z\lesssim1.5$), 
which calls for an extension of such statistical investigations to higher redshifts.
Recently, more complete surveys \citep{2011A&A...526A..30G} have indicated the 
existence of a fraction of GRB host galaxies being more chemically evolved, 
more dusty and of higher mass \citep{2011A&A...534A.108K}. This highlights 
the importance of minimizing optical selection effects in using GRBs as tracers 
of high-redshift star formation and the cosmic star-formation history 
\citep[e.g.,][]{2012ApJ...754...46T,2012ApJ...749...68S,2012ApJ...744...95R,2012A&A...539A.113E,2012A&A...542A.103B,2012ApJ...749L..38T}.

We have conducted a survey of galaxies hosting GRBs, suitable for statistical 
studies and for observations with other existing or future facilities. In 
particular, we have imaged 69 GRB host galaxies to deep limits in $R$ and 
$K_{s}$ and obtained follow-up spectroscopy to determine redshifts and study 
Ly$\alpha$ emission in the higher-redshift systems. The improved redshift 
completeness is also important for statistical studies of GRB properties 
themselves. The details of the survey design and target selection are 
described in Section 2. The observations are described in Section 3, along 
with the catalog of GRB and host galaxy properties, including magnitudes and 
redshifts. Section 4 provides a summary of some of the immediate scientific
results and Section 5 concludes. More detailed discussion and results are 
presented in four companion papers by 
D. Malesani et al.\ (2012, in preparation), 
\citet{2012ApJ...752...62J}, 
\citet{2012arXiv1205.3779M}, and \citet{2012arXiv1205.4036K}.  Supplementary radio, {\it Hubble Space Telescope} ({\hst}), and X-ray 
observations are reported in \citet{2012arXiv1205.4239M}, 
S. Schulze et al.\ (2012, in preparation), and
D. Watson et al.\ (2012, in preparation).

All magnitudes given in this paper are in the Vega system. Throughout our work,
we adopt a cosmology with $\Omega_{\rm m} = 0.27$, $\Omega_\Lambda = 0.73$, 
$H_0 = 71$~km~s$^{-1}$~Mpc$^{-1}$.

\section{Target selection and sample definition \label{selection}}

A statistically meaningful and optically unbiased sample is essential in 
addressing issues such as the GRB redshift distribution, the importance of 
dust obscuration, and the determination of the burst and host luminosity 
functions \citep[see also][]{2012ApJ...749...68S}. We have applied the 
following selection criteria to provide a large and homogeneous sample of 
targets and so as to ensure that selection effects are minimized and well 
understood. The criteria are akin to those introduced by 
\citet{2006A&A...447..897J} and refined by \citet{2009ApJS..185..526F}.

\begin{enumerate}

\item Our starting point is the sample of \swift-detected GRBs. \swift\ is 
the only current facility which can provide a large, homogeneous, 
well-localized sample of GRBs. We pick only events triggered onboard by the
Burst Alert Telescope (BAT) and disregard those (very few) discovered during 
ground analysis.

\item We limit ourselves to long-duration GRBs, and discard all events with 
$T_{90} < 2$~s \citep{1993ApJ...413L.101K}, based on the catalog of 
\citet{2011ApJS..195....2S}.

\item The bursts must have a detected X-ray afterglow, the location of which
is made available less than 12~hr after the trigger. The requirement of the 
existence of an X-ray afterglow effectively makes our sample X-ray selected.
\end{enumerate}

The above criteria are essentially constraints on the nature of the GRBs.
The requirement on the timing in Criterion 3 additionally ensures that 
efficient afterglow searches could be performed. It also works as an effective 
Sun constraint, since those bursts with lines of sight close to the Sun upon
discovery cannot be observed quickly. It should be noted that the fraction
of long GRBs without a detected X-ray afterglow is very low, providing 
the XRT is repointed rapidly
(see below).

Looking for the hosts of all these GRBs would be very demanding. We therefore 
apply further cuts, with the purpose of reducing the sample size in an 
unbiased way. Imposing restrictions 4--6 does not bias the sample toward 
optically bright afterglows; instead each GRB in the sample has favorable 
observing conditions, i.e., useful follow-up observations are likely to be 
secured.

\begin{figure}
\epsscale{1.00}
\plotone{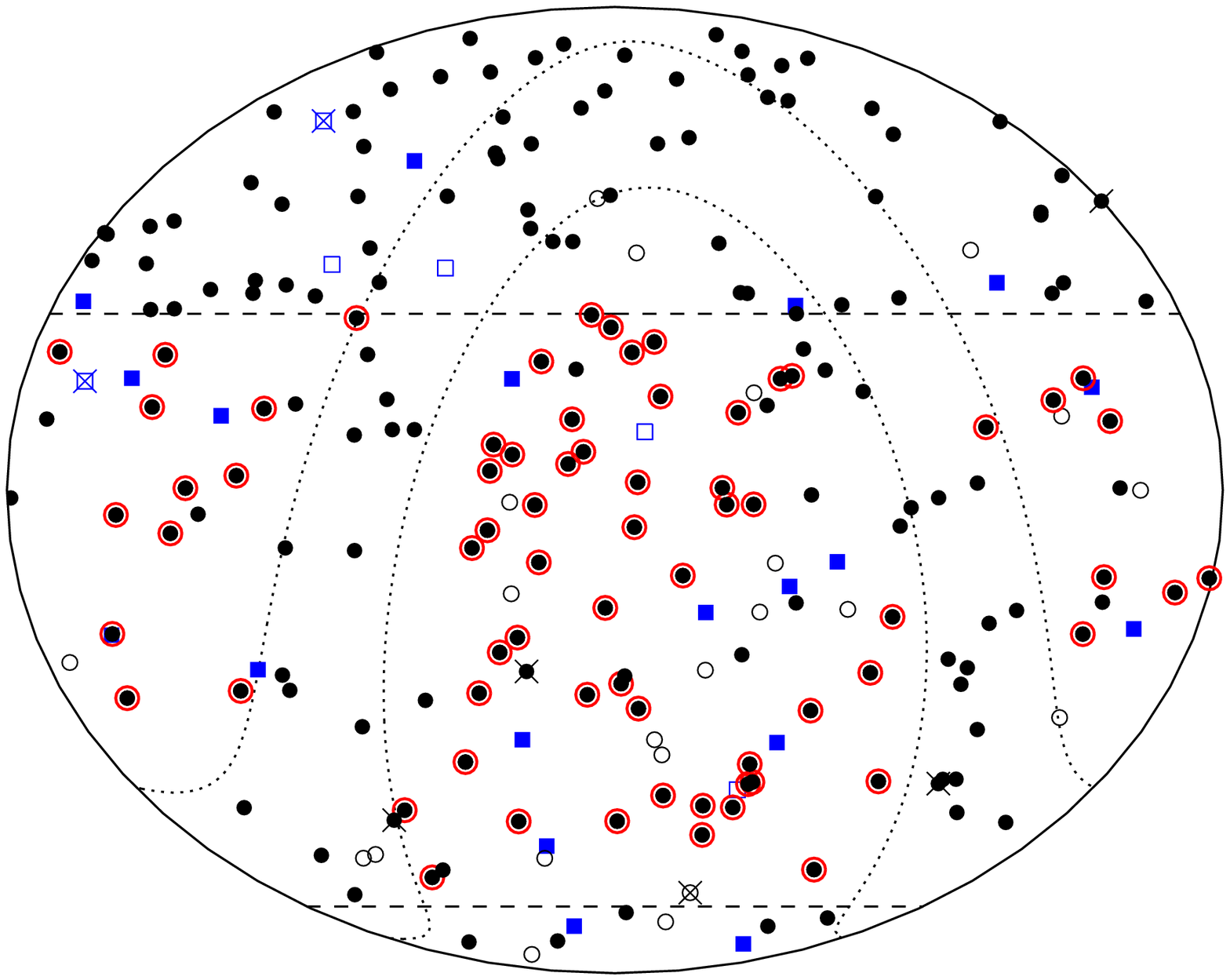}
\caption{All-sky map (Mollweid projection) of the 236 \swift{} GRBs which
occurred between 2005 March 1 and 2007 August 10 (Criterion 7). 
Filled circles: GRBs with \swift/XRT detection. 
Filled circles, encircled in red: GRBs obeying all our sample selection 
Criteria 1--9. 
Empty circles: GRBs with no \swift/XRT detection (excluded by Criterion 3). 
Blue squares: GRBs classified as short (excluded by Criterion 2). 
Crosses: nontriggered GRBs (excluded by Criterion 1). 
The declination cuts are shown with dashed lines ($-70\degr$ and $+27\degr$; 
Criterion 8) and the region with Galactic latitude $|b| < 15\degr$ is
shown with the dotted curves, which roughly corresponds to the sample 
selection criterion $A_V \leqslant 0.5$ mag (Criterion 4). 
\label{f:map}}
\end{figure}

\begin{enumerate}\setcounter{enumi}{3}

\item The Galactic foreground optical extinction as determined by \citet{1998ApJ...500..525S} 
must be $A_V \leqslant 0.5$~mag. 
This makes the detection level almost uniform (i.e., no bias against systems 
that happen to be faint just because of Galactic extinction). As a bonus, all 
fields have high Galactic latitude ($|b| \ga 15\degr$) ensuring that crowding by foreground stars
is not an issue (Figure~\ref{f:map}).

\item As another visibility constraint, we require that the distance on the  sky
between the GRB and the Sun at the time of explosion is more than 55\degr{}
\citep[similar to][]{2006A&A...447..897J}. This is relevant since bursts  closer
to the Sun than this will offer very limited opportunity for ground  follow-up.
In practice, it is not a very demanding constraint because  Criterion 3
effectively enforces a solar angle constraint of this order to satisfy the 
\swift{} Sun-avoidance condition for rapid slews for the narrow-field instruments
(XRT and the Ultraviolet-Optical Telescope, UVOT) of 47\degr.
In fact, during the survey period, only GRB\,070810A had an X-ray afterglow 
announced within 12\,hr but had a discovery line of sight closer to the Sun.

\item We require that there is no nearby star or bright contaminating object
that would render host-galaxy identification and spectroscopy difficult. To 
make this quantitative, we note that, under the typical observing conditions 
of our $R$-band imaging survey, bright stars affect a region of radius
$r \approx 1.8 + 0.4\times\exp[(20-R)/2.05]$~arcsec, where $R$ is the USNO-B
magnitude \citep{2003AJ....125..984M}. This relation is valid in the range 
$11 \la R \la 19$. We thus require that no star with $R < 19$ is located 
on the CCD detected within this distance from the X-ray error circle 
edge. In two cases, GRBs\,060604 and 070721B, very bright stars ($R < 10$) at 
distances $\ga 1\arcmin$ would nominally violate the above limit, but their 
effect was minimized by offsetting the pointing such that they were located 
outside of the detector.
\end{enumerate}

The aforementioned six criteria are applicable in principle to any survey of
\swift{} GRB positions. Our survey was carried out with the 
VLT, and was restricted to bursts detected during a specific 2.5 yr
period, so that for this sample we applied two additional criteria.

\begin{enumerate}\setcounter{enumi}{6}
\item Bursts triggered between 2005 March 1 and 2007 August 10. During this 
period \swift{} was fully operational and automatic slews were routinely 
enabled. There were 236 GRBs (229 triggered) in this 
time.\footnote{We do not consider GRBs 060204C, 060602B, and 070610, which were
very likely due  to Galactic sources 
\citep{2006GCN..4679....1S,2006GCN..5208....1P,2008ApJ...678.1127K,2008Natur.455..506C}.} 

\item In order to be well placed for observation with the VLT we limit the
declination range to be $-70\degr < \delta < +27\degr$ (J2000.0). 

\end{enumerate}

These criteria can easily be extended
by using other observatories and different points in time.

After enforcing these criteria, we end up with a sample of 69 GRBs. The advantage of 
our selection is illustrated by the large success of afterglow follow-up for
bursts in this subset. Based on Jochen Greiner's Web 
site,\footnote{\texttt{http://www.mpe.mpg.de/$\sim$jcg/grbgen.html}} 
optical/NIR afterglows have been discovered for 75\% (52/69) of the bursts
in our sample, compared to 66\% for the overall \swift{} sample. More 
importantly, spectroscopic redshifts are available for 55\% of our targets 
(38 GRBs) compared to only 37\% in the overall \swift{} sample.

Our final criterion concerns the localization accuracy of the GRBs.

\begin{enumerate}\setcounter{enumi}{8} 
\item The X-ray localization uncertainty must be better than or equal to 
2~arcsec (90\% error radius).
%
%
This constraint is dictated by the necessity to perform meaningful host
searches. Larger error boxes would lead to too many ambiguous detections. For
bursts without an optical/NIR localization, we adopt the best constrained
position among the catalogs provided by \citet{2007AJ....133.1027B} and
\citet{2009MNRAS.397.1177E}. In principle, a more accurate localization could 
be available from optical or radio data, but we only base our selection on the 
X-ray position, to retain a dependence solely on high-energy properties.
\end{enumerate} 

We summarize the selection criteria in Table~1. Our final sample consists
of 69 GRBs and is presented in 
Table~2.
Figure~\ref{f:map} shows the distribution in the sky of all GRBs detected by 
\swift{} during the chosen time period with the TOUGH sample explicitly marked.

\begin{figure}
\epsscale{1.00}
\plotone{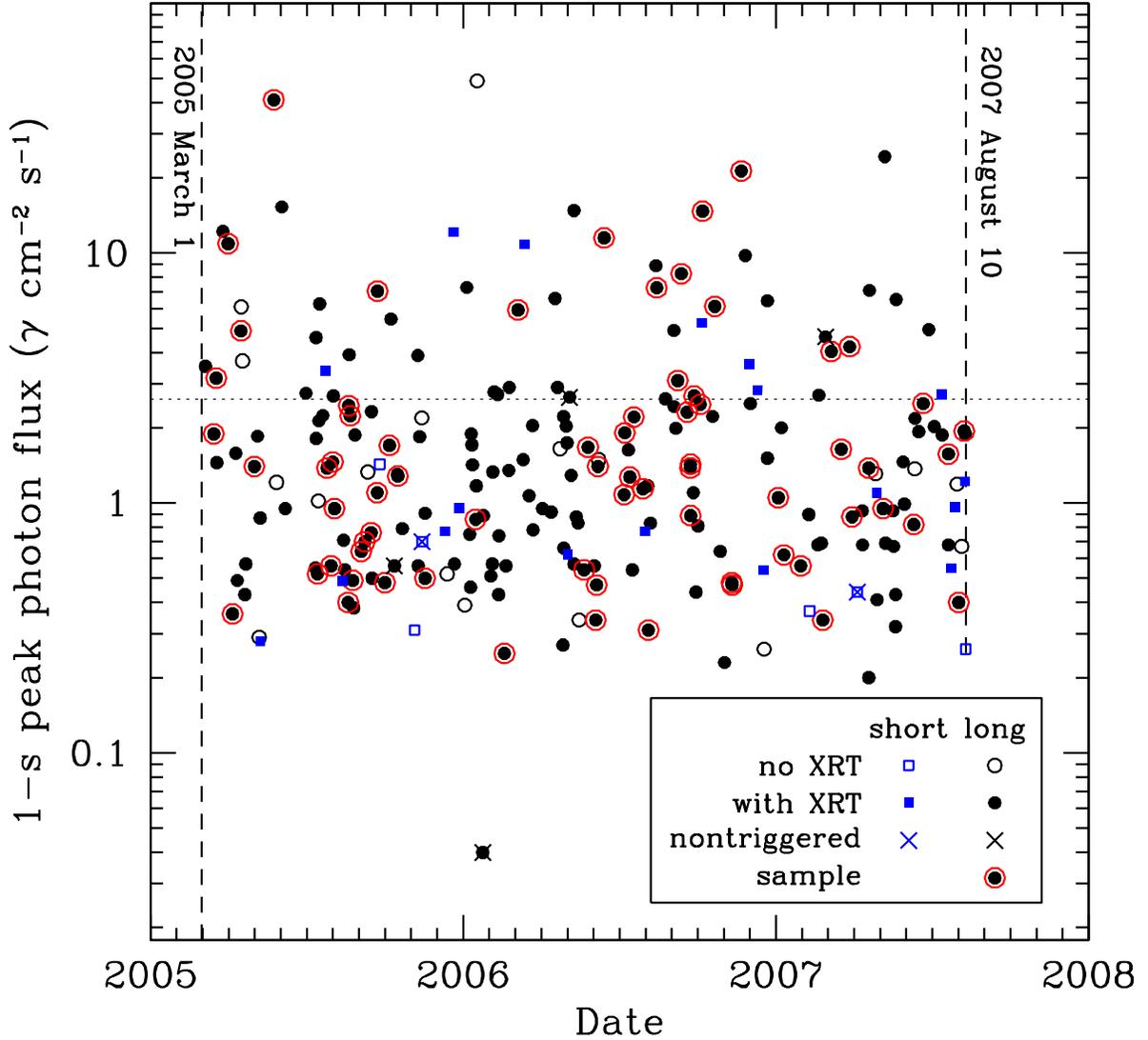}
\caption{1-s \swift/BAT peak photon flux (15--150 keV) for all \swift{} GRBs observed
during the time span of the TOUGH survey. Empty
circles: long  bursts without an X-ray afterglow. Filled circles: long GRBs with
\swift/XRT  detection. Filled, encircled circles: GRBs obeying all our sample
selection  Criteria 1--9. Squares: GRBs classified as short ($T_{90}<2$~s).
Crosses: nontriggered GRBs. There seems to be a constant and uniform detection
level as a function of time \citep[the very faint system at 0.04
$\gamma$ cm$^{-2}$ s$^{-1}$ is the unusual
GRB 060123;][]{2006GCN..4608....1C}.
The vertical dashed lines bracket the selected
time span for the sample. The horizontal dotted line is the peak flux limit adopted by
\citet{2012ApJ...749...68S}.\label{f:pf_time}}
\end{figure}

Below we discuss the effects of the various selections. Figure~\ref{f:pf_time} 
shows the 1~s BAT 15--150 keV peak photon flux (from \citealt{2011ApJS..195....2S}) versus the
date.  As can be  seen, our targets uniformly sample the full BAT distribution
and there is no  obvious long-term change in the detection threshold as a
function of time,  which seems to be consistently at a level of about
$0.4~\gamma$~cm$^{-2}$~s$^{-1}$ (the  faintest target GRB\,060218 is detected at
$0.25~\gamma$~cm$^{-2}$~s$^{-1}$).  This shows that we are starting from a
stable, well-controlled set of events  (Criterion 1), although we note that the
effective trigger threshold does vary  somewhat with many factors, such as the
current background level, the position  of the burst within the BAT field, and
also the form of the particular GRB  light curve (for example, the detection of
some bursts is only made after  integrating up more than the 1 s binning we
consider here). Out of the 69  TOUGH bursts, 56 (13) were rate (image)
triggers. 

Our Criterion 3 is the detection of an X-ray afterglow. During the considered 
period, only 25 triggered GRBs were not detected (or observed) by XRT (11\% of
the overall \swift{} sample).
Restricting to long-duration bursts (Criterion 2) with prompt slew (less than
3600\,s), however, we find that only 5 events out of 180 (2.8\%) were not
detected by XRT. This shows that, in practice, XRT selection has a negligible 
impact on the size of our sample, although it does  mean that rare GRBs with 
very faint X-ray afterglows will be systematically underrepresented.

\begin{figure}
\epsscale{1.00}
\plotone{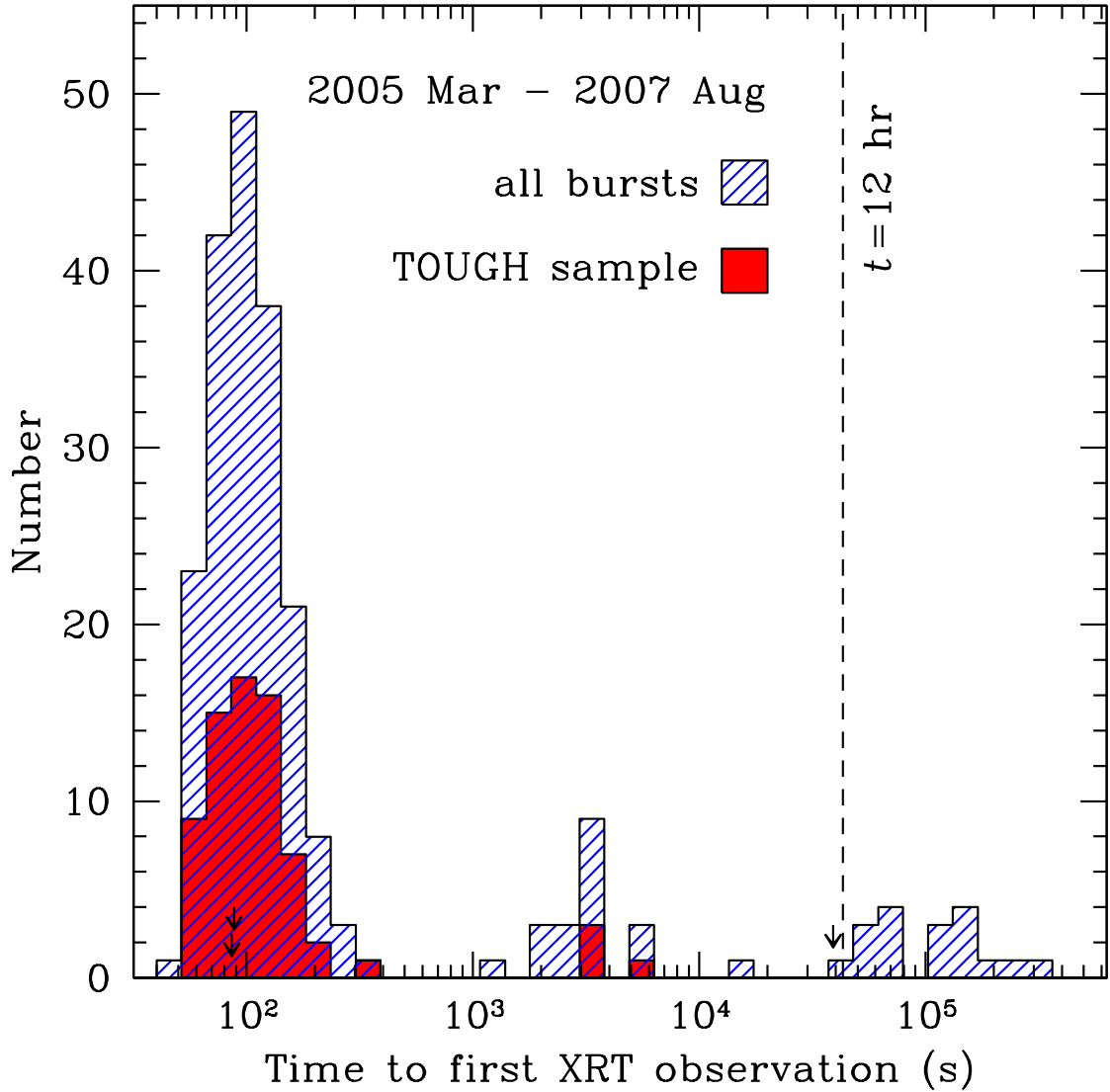}
\caption{Distribution of times post-trigger to the first XRT observation for 
all \swift{} GRBs happening between 2005 March 1 and 2007 August 10. For 
12 bursts, the XRT was not repointed. The fraction constituting the 
TOUGH sample is indicated in red. The  vertical dashed line
marks the sample limit $t \leqslant 12$\,hr, which is fulfilled by 92\% of the
repointed bursts (207/224).  Three GRBs (050603, 060223A, and 070810A) have
$t_{\rm XRT} < 12$~hr but the notice came later than 12~hr. These GRBs are
indicated by arrows in the histogram.\label{f:tXRT_histo}}
\end{figure}

\begin{figure}
\epsscale{1.00}
\plotone{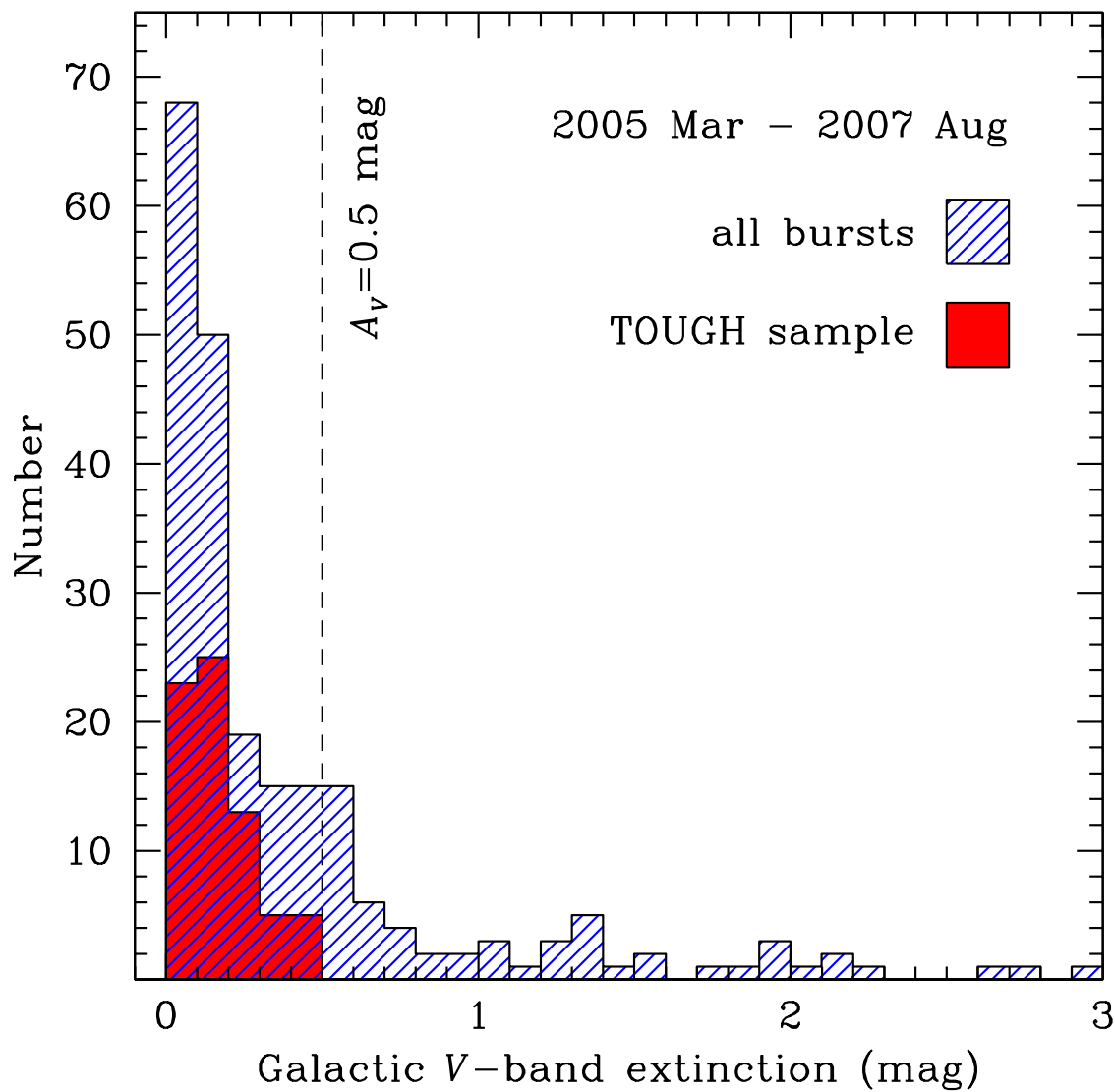}
\caption{Distribution of Galactic foreground extinction for all \swift{} 
GRBs detected
between 2005 March 1 and 2007 August 10 (13 bursts have 3 mag $< A_V < 17$~mag and
are not shown in the figure). The fraction constituting the TOUGH sample is 
indicated in red. The vertical
dashed line marks the sample limit $A_V \leqslant 0.5$~mag. 
Seventy-one percent of all bursts
(167/236) have $A_V \leqslant 0.5$~mag. \label{f:AV_histo}}
\end{figure}

Figures~\ref{f:tXRT_histo} and \ref{f:AV_histo} illustrate the distributions 
of the times to first XRT observation (taken from the \swift{} GRB
table\footnote{\texttt{http://swift.gsfc.nasa.gov/docs/swift/archive/grb\_table/}};
akin to but not identical to Criterion 3) and the Galactic $V$-band 
extinction \citep[][Criterion 4]{1998ApJ...500..525S} for the TOUGH sample
as well as the full distribution of GRBs.

\begin{figure}
\epsscale{1.00}
\plotone{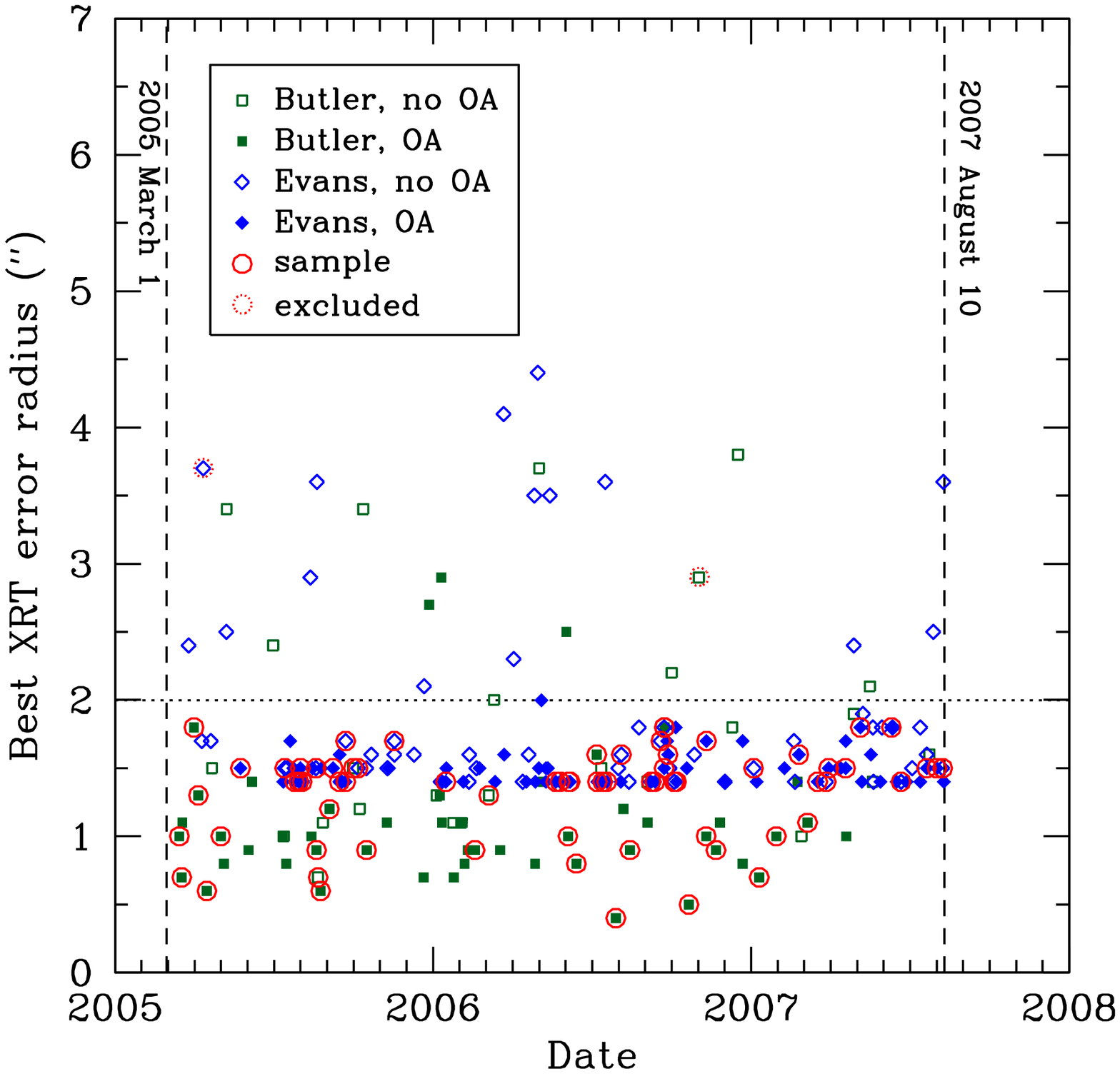}
\caption{\swift/XRT error radius (90\% encircled likelihood)
as a function of date for GRBs detected during the time span of the 
TOUGH survey. Filled and empty points
indicate bursts with and without a detected optical/NIR afterglow (OA), respectively.
Green squares indicate positions from the \citet{2007AJ....133.1027B} catalog, 
while blue diamonds are from the UVOT-enhanced list 
\citep{2007A&A...476.1401G,2009MNRAS.397.1177E}. Our sample targets are circled
in red. During the period 2005 March 1 through  2007 August 10 (marked by
vertical dashed lines), 89\% of the GRBs have an XRT  error circle radius
$\leqslant 2\arcsec$ (horizontal dotted line). The two dotted circles indicate
the two bursts that are excluded only by Criterion 9 ($r \leqslant 2\arcsec$;
see Figure~\ref{f:error_flux}). The average  error radius is
1\farcs7.\label{f:error_time}
}
\end{figure}

\begin{figure}
\epsscale{1.00}
\plotone{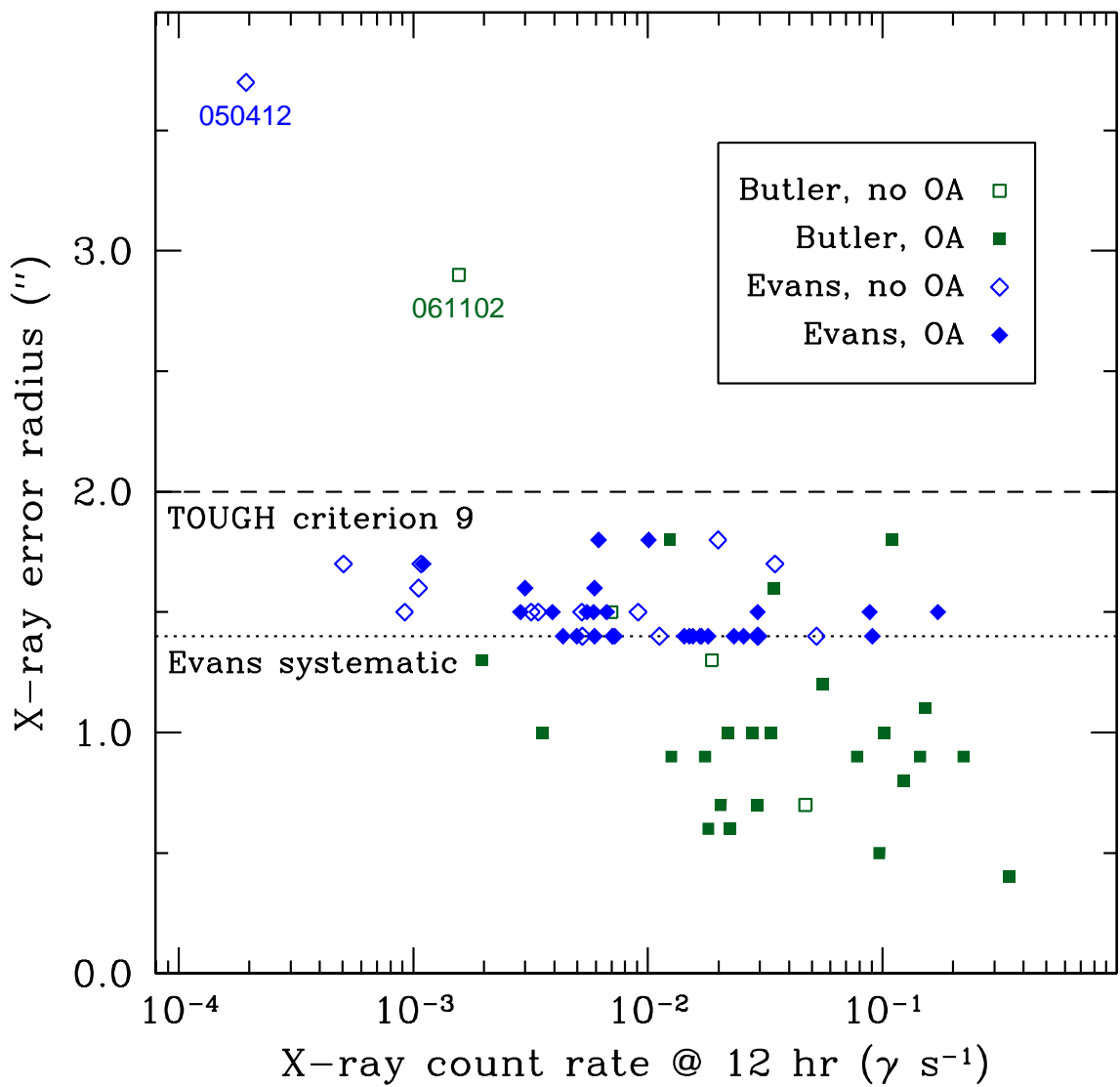}
\caption{X-ray error radius vs.\ the \swift/XRT count rate at 12~hr (0.3--10
keV). Only bursts which ``survive'' Criteria 1--8 are shown. 
Filled and empty points indicate bursts with and without a detected 
optical/NIR afterglow (OA), respectively.
The horizontal
dashed line indicates the TOUGH constraint set by Criterion 9. Notice the flat
distribution, which implies that we only minimally bias our sample. Indeed,
only two GRBs are excluded. The dotted line indicates the systematic error
contributing to the UVOT-enhanced uncertainties \citep{2009MNRAS.397.1177E}.
\label{f:error_flux}}
\end{figure}

Criterion 9 introduces a potential bias in our requirement of an accurate XRT
position, which might lead us to favor optically bright afterglows. The accuracy
of the X-ray position is influenced by many factors, several of which are not
directly related to the GRB properties. For example, the
astrometrically corrected positions \citep{2007AJ....133.1027B} require the
detection of serendipitous X-ray sources in the field other than the GRB. This
depends more on the total exposure time than on the afterglow brightness.
Figure~\ref{f:error_time} shows that the accuracy of X-ray positions has been
stable during the considered time span. In fact, many of the UVOT-enhanced
positions are dominated by the systematic error term which is independent of the
afterglow brightness. There is a possible dearth of very accurate positions
after 2007 March, which may be due to shallower exposures following a change of
the \swift{} pointing strategy. In Figure~\ref{f:error_flux}, we investigate the
impact of rejecting bursts without an accurate localization. We pick all GRBs
which would pass selection criteria 1--8, regardless of the X-ray accuracy. We
then plot their 90\% error radii \citep{2007ApJ...656.1001B,2009MNRAS.397.1177E}
as a function of the count rate at 12~hr, obtained by interpolating or
extrapolating the light curves posted
online\footnote{\texttt{http://www.swift.ac.uk/xrt\_curves/}} \citep{2009MNRAS.397.1177E}.
Figure~\ref{f:error_flux} shows that, indeed, faint afterglows have on the average
larger error radii, but the effect is very small (a factor of $\approx 3$ across
more than two orders of magnitude in the count rate). Thus, among 71 GRBs
considered here, only 2 are discarded (less than 3\%), showing that the impact
on our sample is limited.

\section{Survey and catalogs\label{goals}}

Our observational program consists of deep $R$ and $K_{s}$ imaging and 
spectroscopy obtained with the FORS1, FORS2, ISAAC, and X-shooter instruments 
at the ESO VLT. All VLT observations were conducted at least 50 days after the 
GRB explosion in order to avoid possible contamination from the afterglow or an
associated supernova. The possible afterglow contamination in our images is
quantified in 
D. Malesani et al.\ (2012, in preparation). 
Here, we briefly describe the overall results, as
summarized in 
Table~3.
Detailed survey results are reported in four accompanying papers. Results on 
individual targets or preliminary results are reported in
\citet{2007ApJ...669....1R},
\citet{2007GCN..6997....1J,2009AIPC.1133..455J,2008mgm..conf.2019J,2011AdSpR..47.1416J,2011AN....332..276J,2011AIPC.1358..265J},
\citet{2008MNRAS.388.1743T},
\citet{2008ApJ...676.1151T,2010A&A...523A..70T},
\citet{2008ApJ...681..453J},
\citet{2009ApJ...696..971X},
\citet{2009ApJ...697.1725E},
\citet{2009A&A...497..729F},
\citet{2009ApJS..185..526F,2008mgm..conf..726F},
\citet{2009AIPC.1111..513M,Malesani_Venice},
\citet{2012MNRAS.421...25S}, and
\citet{2012arXiv1207.6088S}.
In Tables~2 and 3,
we also list 10 extra hosts that were observed for our program, but that do not
obey all of the TOUGH sample criteria.

\subsection{Optical imaging\label{r}}

For the purposes of identification of hosts and brightness estimation we have 
obtained deep images of the targets in the $R$ band (typically rest-frame UV)  
to magnitude limits in the range $R = 26\mbox{--}28$ mag. The seeing FWHM of 
the $R$-band images ranges from 0\farcs50 to 1\farcs26 with a median of 
0\farcs71. D. Malesani et al.\ (2012, in preparation) discuss our
procedure for identifying a host galaxy, including a quantification of the 
chance probability of wrongly identifying an unrelated galaxy in the error 
circle as the host. We note that there is always a small risk of wrong host 
identification in both imaging and spectroscopy. While this could lead to 
erroneous conclusions when discussing individual systems, 
it will have 
little effect on statistical results inferred from a sufficiently large 
sample, like the one presented here, as long as the probability for false 
identification is low. 

\begin{figure}
\epsscale{1.00}
\plotone{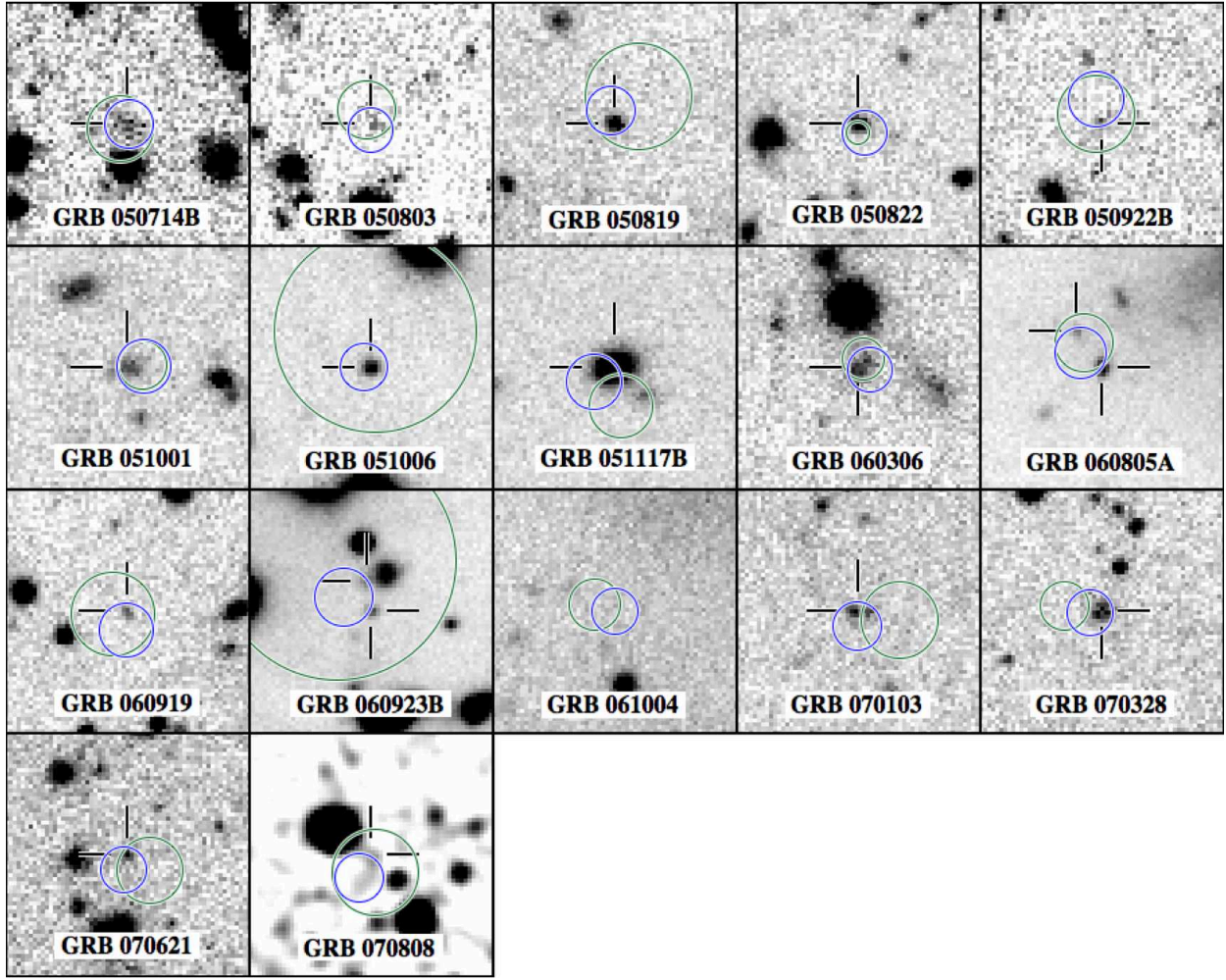}
\caption{
Mosaic of the fields of 17 XRT-localized GRBs with no optical/NIR afterglow. Each 
panel, showing the $R$-band image, is $15\arcsec \times 15\arcsec$ in size 
(north is up and east is left). Blue and green circles identify the XRT positions 
by \citet{2009MNRAS.397.1177E} and \citet{2007AJ....133.1027B}, respectively.
\label{f:malesani1}}
\end{figure}

\begin{figure}
\epsscale{0.58}
\plotone{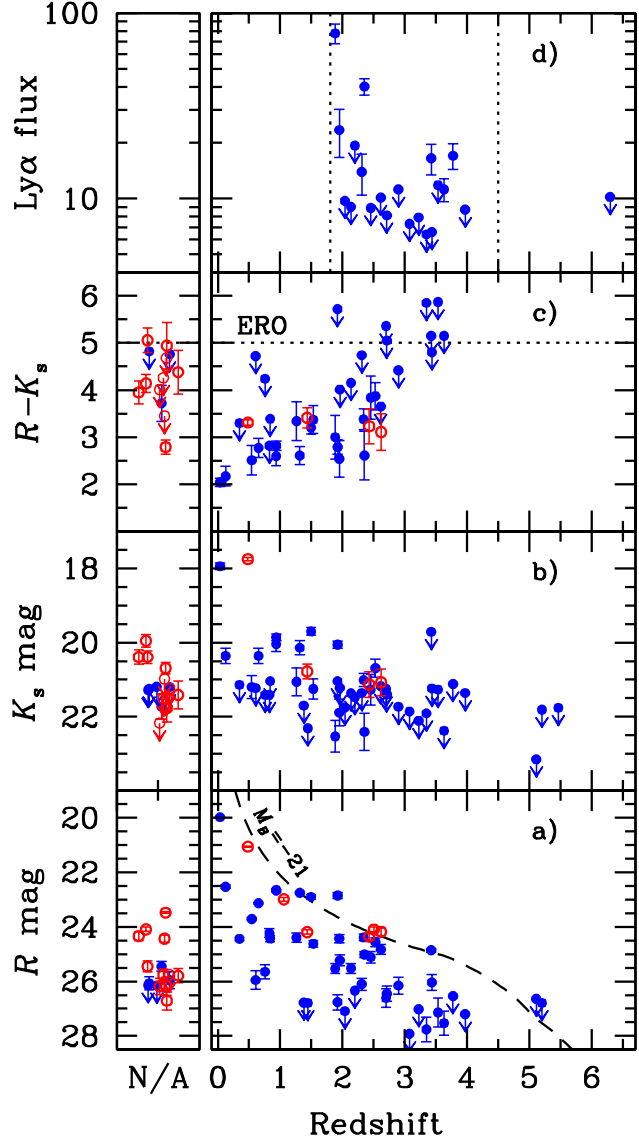}
\caption{TOUGH sample properties as a function of redshift. Optically localized
systems are shown as blue filled circles, XRT localized systems are shown as
open red circles. Upper limits are shown as arrows. Hosts without a reported
redshift are plotted on the left side of the diagram.  (a) $R$-band host
magnitudes (Vega system). 
The dashed curve corresponds to the observed $R$-band magnitude for an 
$M_B = -21$ (Vega) object assuming $F_{\nu} \propto \nu^{-0.5}$ and 
IGM opacity as described by \citet{2008A&A...491..465D}.
(b)
$K_{s}$-band host magnitudes (Vega system). (c) $R-K_{s}$ colors. The ERO limit
of $R-K_{s}=5$ is shown as a dotted line. Two EROs are detected. (d) 
Ly$\alpha$ detections and upper limits from \citet{2012arXiv1205.3779M}. The upper
limit reported for GRB 050904 at $z=6.30$ \citep{2006PASJ...58..485T} was 
adjusted to be comparable with those reported in \citet{2012arXiv1205.3779M}. The 
Ly$\alpha$ fluxes are in units of $10^{-18}$ erg~cm$^{-2}$~s$^{-1}$. The 
vertical dotted lines mark the redshift interval where Ly$\alpha$ was
searched in our data.\label{f:malesani2}}
\end{figure}

We have identified host galaxies for 55 systems, corresponding to 80\%
of the sample. The coordinates of the identified host galaxies are reported in 
Table~3.
In Figure~\ref{f:malesani1}, we show a mosaic of GRB host galaxies that are 
only localized by the XRT (a mosaic of all identified host galaxies is 
presented in D. Malesani et al.\ (2012, in preparation). 
These 17 galaxies (25\% of our sample) are 
typically missed from other surveys.
The observed magnitudes are shown in Figure~\ref{f:malesani2}. 
It is evident that host galaxies are generally sub-luminous, in the sense of 
being fainter than $L^*$ at $z=0$. We note that the simple detection of a host 
galaxy in the $R$ band can be used to set an upper limit to the redshift of 
the GRB of $z \la 5.6$ \citep{2007ApJ...669....1R,2012ApJ...752...62J}.

\subsection{Near-infrared imaging\label{k}}

To obtain basic color information for the targets we also obtained 
$K_{s}$-band 
imaging (typically rest-frame visual) to a limiting magnitude of about
22. The seeing FWHM of the $K_{s}$-band images ranges from 0\farcs34 to 
0\farcs83 with a median of 0\farcs51. These data allow detection of 
extremely red objects (EROs, $R-K_{s} > 5$) for all 
$R$-band-detected targets. Although the majority of previously studied GRB 
hosts have had blue optical--IR colors, a small number of ERO hosts have been 
identified \citep[e.g.,][]{2006ApJ...647..471L,2007ApJ...660..504B,2011ApJ...736L..36H,2012MNRAS.421...25S,2012arXiv1202.1434R}. The resulting $R-K_{s}$ colors 
for our sample are shown in Figure~\ref{f:malesani2}. These are roughly 
consistent with \citet{2003A&A...400..499L}. All host galaxies detected in 
$K_{s}$ were also detected in $R$. We have identified two EROs among our 
sample, the host galaxies of GRB\,050714B and GRB\,070808 which both have 
$R-K_{s} \approx 5$.

\subsection{Spectroscopy\label{s}}

The identification of the hosts
was used to determine their suitability 
for spectroscopic follow-up. For targets with no afterglow absorption redshift 
that are sufficiently bright for spectroscopic redshift determination (host 
galaxies generally are emission-line galaxies) we have obtained spectra with 
the purpose of making the redshift determinations for our sample as complete 
as possible. This is a crucial step in making any statistical inferences about 
the redshift distribution of GRBs and their host galaxies, in particular with 
reference to the star formation history of the universe. However, we note that 
the selection of galaxies at this stage based on their apparent $R$-band
magnitudes could introduce an optical bias into our sample, and that securing 
redshifts for the remaining (fainter) galaxies remains a priority. As reported 
by \citet{2012ApJ...752...62J} and \citet{2012arXiv1205.4036K}, we have obtained spectra of 23 host 
galaxies in the TOUGH sample for which we have determined 15 new redshifts 
(and, unfortunately, found that 3 previously reported redshifts of the sample
were erroneous). In total, we now have redshifts for 53 galaxies out of 69 
in the sample. Moreover, even in cases where spectroscopy was unsuccessful in 
determining a redshift, spectra and imaging were used to set useful 
constraints, e.g., from an upper limit of the position of the Lyman limit or 
Ly$\alpha$ breaks or from the absence of emission lines. 

\section{Discussion\label{d}}

We have verified that there are no obvious trends in the redshifts or
magnitudes of the TOUGH host galaxies as a function of GRB epoch.

Below we summarize some of the results of the TOUGH survey, with special
emphasis on the implications of having compiled an X-ray selected sample
of GRB host galaxies, and obtaining photometry and spectroscopy of a
sample of GRB host galaxies extending to faint and high-redshift systems.

\subsection{Redshift distribution\label{z}}

\begin{figure}
\epsscale{1.00}
\plotone{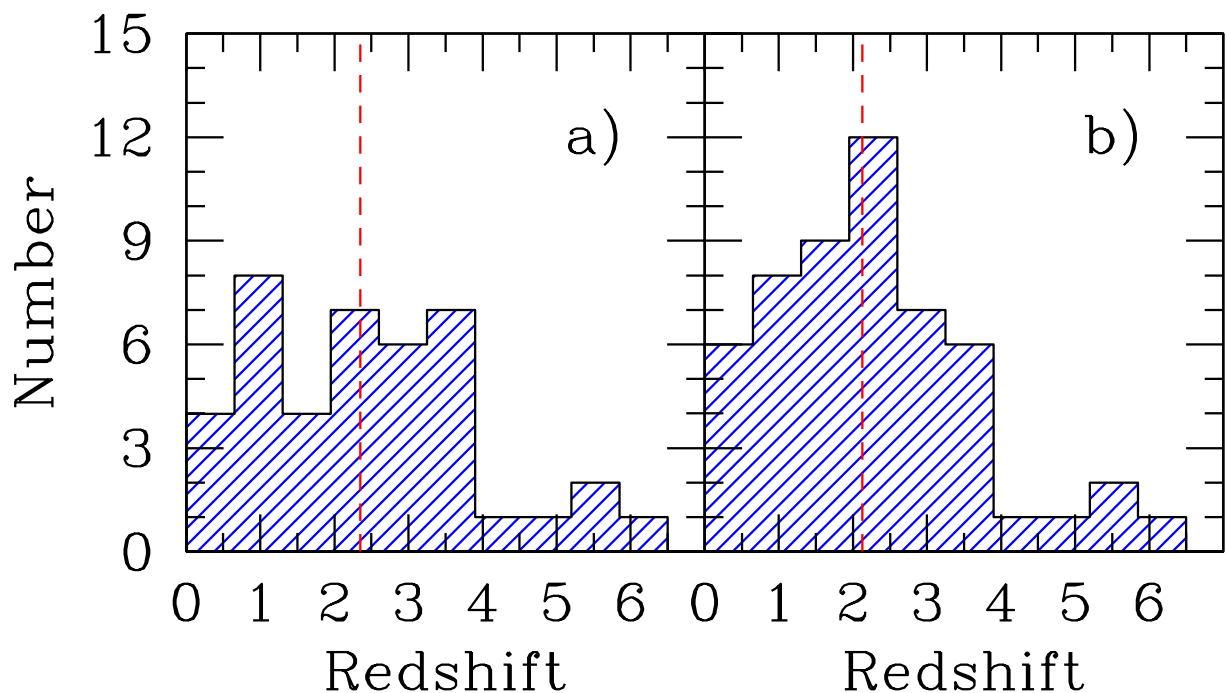}

\caption{Redshift distribution of the TOUGH sample. In both panels the 
dashed vertical line indicates the median redshift. (a) Without 
considering the TOUGH observations, 38 redshifts were considered secure with 
a median redshift of $z =  2.35$. (b) Our TOUGH 
spectroscopic observations add 15 new redshifts and demonstrate that three 
redshifts reported in the literature are erroneous \citep{2012ApJ...752...62J}. 
Here the median redshift is $z = 2.14$. 
\label{f:zdistribution}}

\end{figure}

In Figure~\ref{f:zdistribution}, we show the redshift distribution prior to our 
survey and compare to the redshift distribution resulting from our survey.  
All redshifts are given in 
Table~3.
The new redshift distribution is more smooth with a conspicuous peak at around 
redshift 2 with a median redshift of $2.14\pm0.18$. The median redshift of
the entire TOUGH sample (i.e., including systems with no redshift constraint)
is robustly determined to be between 1.5 and 2.5
(Table~4). As discussed by \citet{2012ApJ...752...62J} our redshifts and redshift 
constraints allow us to 
distinguish models for the GRB rate and 
luminosity function and formation biases. 

\begin{figure}
\epsscale{1.00}
\plotone{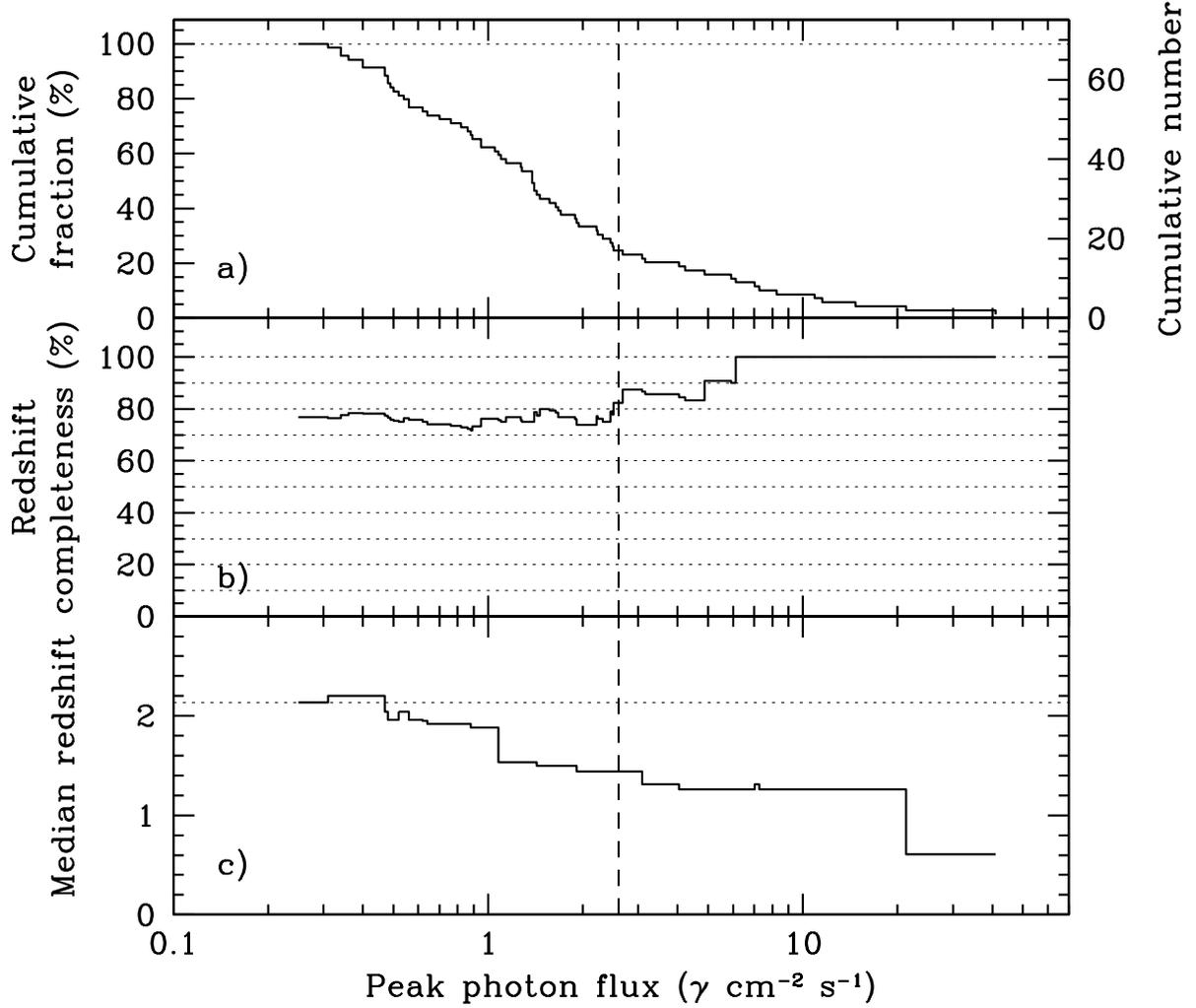}
\caption{
Redshift and BAT 15--150 keV peak photon flux. The vertical dashed line 
indicates the peak photon flux limit adopted by 
\citet{2012ApJ...749...68S}. 
(a) 
The cumulative distribution of peak photon fluxes of the TOUGH bursts. 
(b) 
The redshift completeness fraction of the TOUGH sample bursts as a function 
of the limiting peak photon flux. The redshift completeness is constant 
around 75\% below $\approx 3~\gamma$ cm$^{-2}$ s$^{-1}$. In order to
significantly increase the completeness fraction, one would have to discard 
at least $> 75\%$ of all bursts. 
(c) 
The median 
redshift of TOUGH GRBs with known redshift
brighter than the corresponding peak flux. Fainter GRBs are on average at
higher redshift.
\label{f:completeness}}
\end{figure}

To illustrate the importance of sample selection criteria for the determination 
of the median redshift for GRBs we plot the redshift completeness as a 
function of BAT (15--150\,keV) peak photon flux in Figure~\ref{f:completeness}. 
The median 1 s peak flux for the TOUGH sample is 
$1.4~\gamma$~cm$^{-2}$~s$^{-1}$ (Figure~\ref{f:completeness}). Subsamples with 
or without redshift give a consistent value for the median 1 s peak flux, as 
do subsamples with or without an optical/NIR afterglow. About 75\% of the GRBs in 
our sample have lower peak photon flux than the cutoff adopted 
by \citet{2012ApJ...749...68S} ($\ge 2.6~\gamma$~cm$^{-2}$~s$^{-1}$). Our 
redshift completeness is in the range 70\%--80\% below that value and reaches 
almost 90\% slightly above. The median redshift increases almost monotonically 
with lower peak flux, from 1.38 at a cutoff at $2.6~\gamma$~cm$^{-2}$~s$^{-1}$ 
to 2.12 for a cutoff at $0.4~\gamma$~cm$^{-2}$~s$^{-1}$. Tying to the 
BATSE luminosity function, \citet{2012ApJ...749...68S} modeled their results 
and predicted a median redshift of $z = 2.05\pm0.15$ to that limit, consistent 
with our value. 

\begin{figure}
\epsscale{1.00}
\plotone{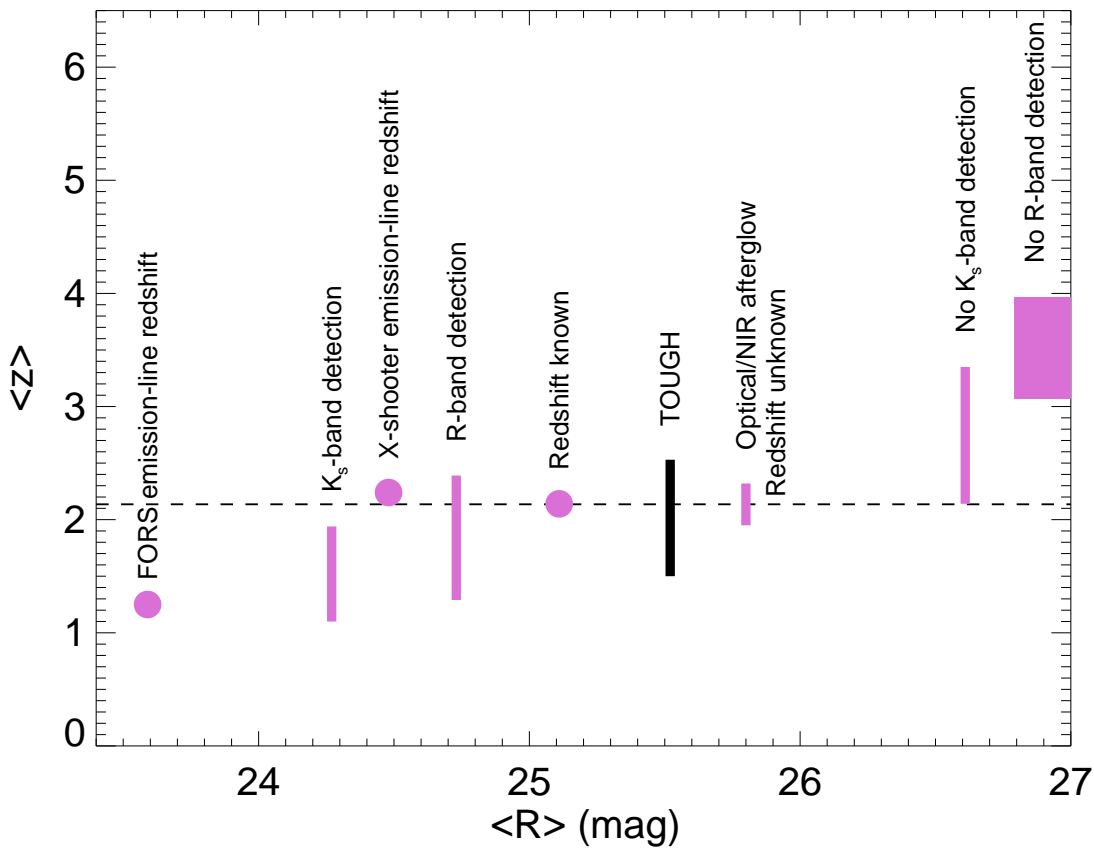}
\caption{
Median redshift (with a range of uncertainty where the median is not uniquely 
determined) against median $R$-band host magnitude for various subsets of 
our sample. In the case of those without an $R$-band detection, the set has 
been placed at $R\approx27$.
The figure broadly illustrates that the fainter hosts are typically at
higher redshift, characteristic of our magnitude limit.
The subset of hosts without redshift has a median $R\approx25.9$ which
suggests a median redshift above 2 if it follows the same trend.
\label{f:Rz}}
\end{figure}

However, obtaining redshifts of fainter galaxies may shift the median 
redshift somewhat higher. As indicated by \citet{2012arXiv1205.4036K} the faint host 
galaxies targeted with X-shooter have been found at significantly higher 
redshifts compared to brighter targets. We therefore suspect that the remaining 
very faint host galaxies may lead to even higher average redshifts. 
Figure~\ref{f:Rz} shows the median redshift as a function of the median 
$R$-band magnitude of the hosts (the corresponding values are given in 
Table~4). The median redshift of the TOUGH sub-samples is lowest for the 
bright hosts for which the redshift is determined from emission lines using 
FORS, higher for fainter hosts where the redshift is determined using 
X-shooter, and higher still for bursts where the host remains undetected.
$K_{s}$-detected host galaxies are at significantly lower redshifts than
galaxies which are not detected in $K_{s}$. The overall trend seems to be
consistent with the expectation that fainter GRB host galaxies are on
average at higher redshifts.

TOUGH for the first time picks up high-redshift GRB hosts in significant 
numbers. The compilations of \citet{2009ApJ...691..152C} and 
\citet{2009ApJ...691..182S} count eight optically detected GRB host galaxies at 
$z > 2$ of which two are \swift\ GRBs. Table 3 lists a total of 18 
TOUGH-detected GRB host galaxies at $z > 2$, of which 17 are new, as well as 
one 
non-TOUGH $z > 2$ GRB host galaxy.  Such a large sample opens the field for 
more detailed studies of individual systems and of high-redshift GRB hosts 
as a whole (see also Section~\ref{sf}).

\subsection{Galaxy colors and magnitudes\label{rk}}

Of special interest is the study of dark bursts which are believed to be 
caused mostly by dust obscuration \citep[e.g.,][]{2001ApJ...562..654D,
2007ApJ...669.1098R,2008ApJ...681..453J,2008MNRAS.388.1743T,
2009AJ....138.1690P,2011A&A...534A.108K,2012MNRAS.421...25S,2012arXiv1202.1434R}. Hence, it is expected that the 
properties of the host galaxies of XRT-only GRBs may be different from those 
hosting GRBs with detected optical/NIR afterglows. 

\begin{figure}
\epsscale{0.5}
\plotone{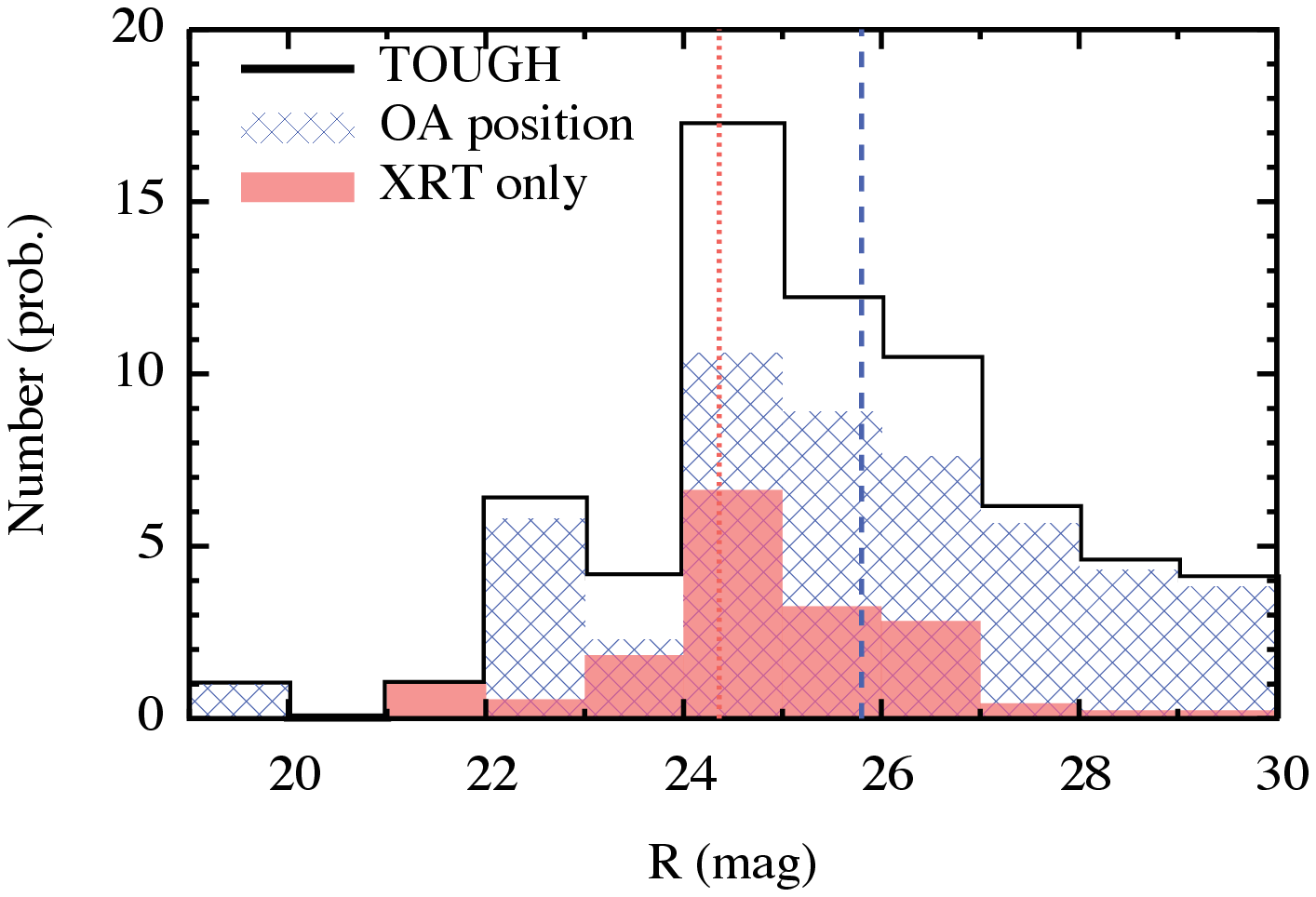}
\plotone{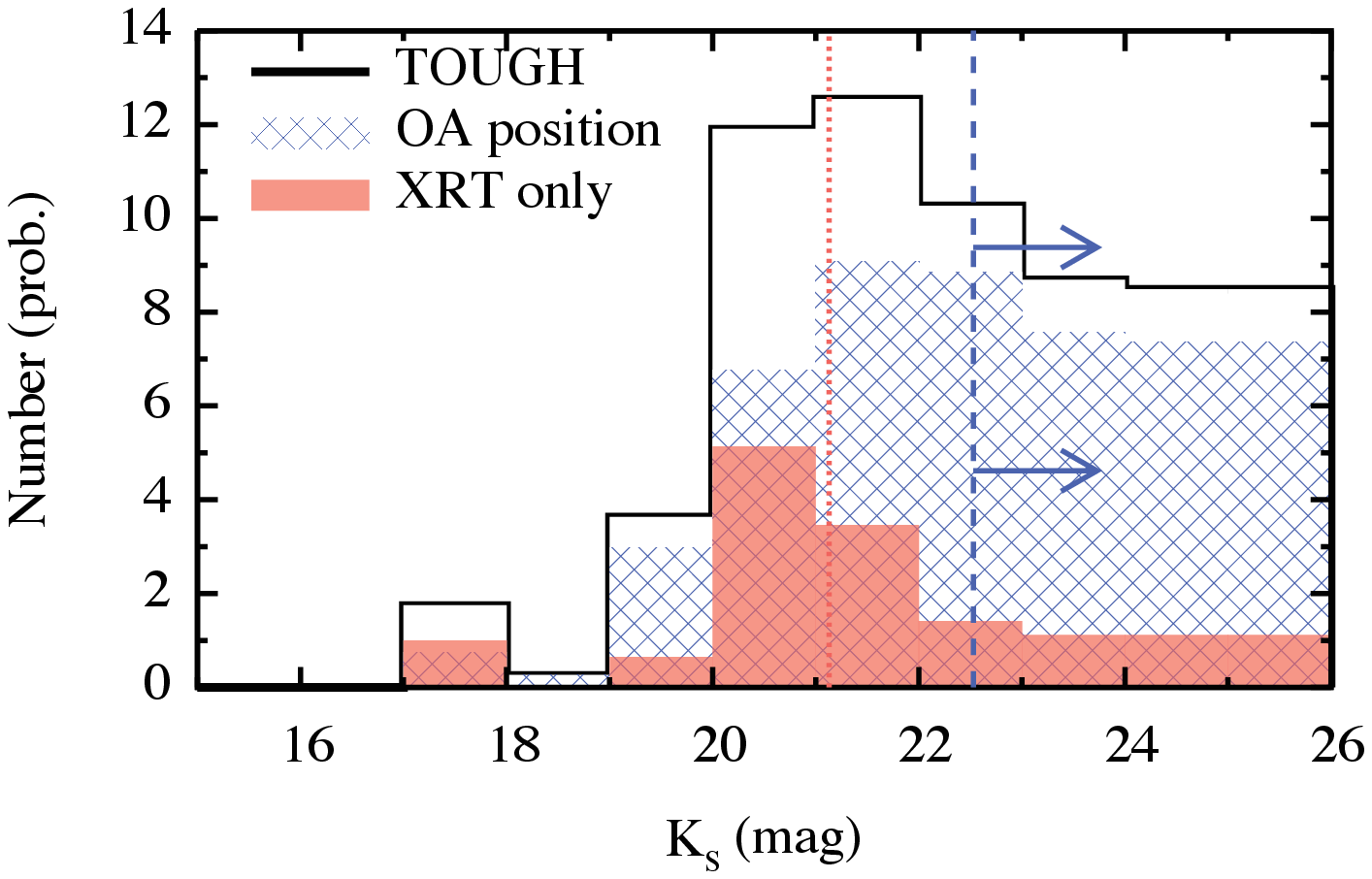}
\plotone{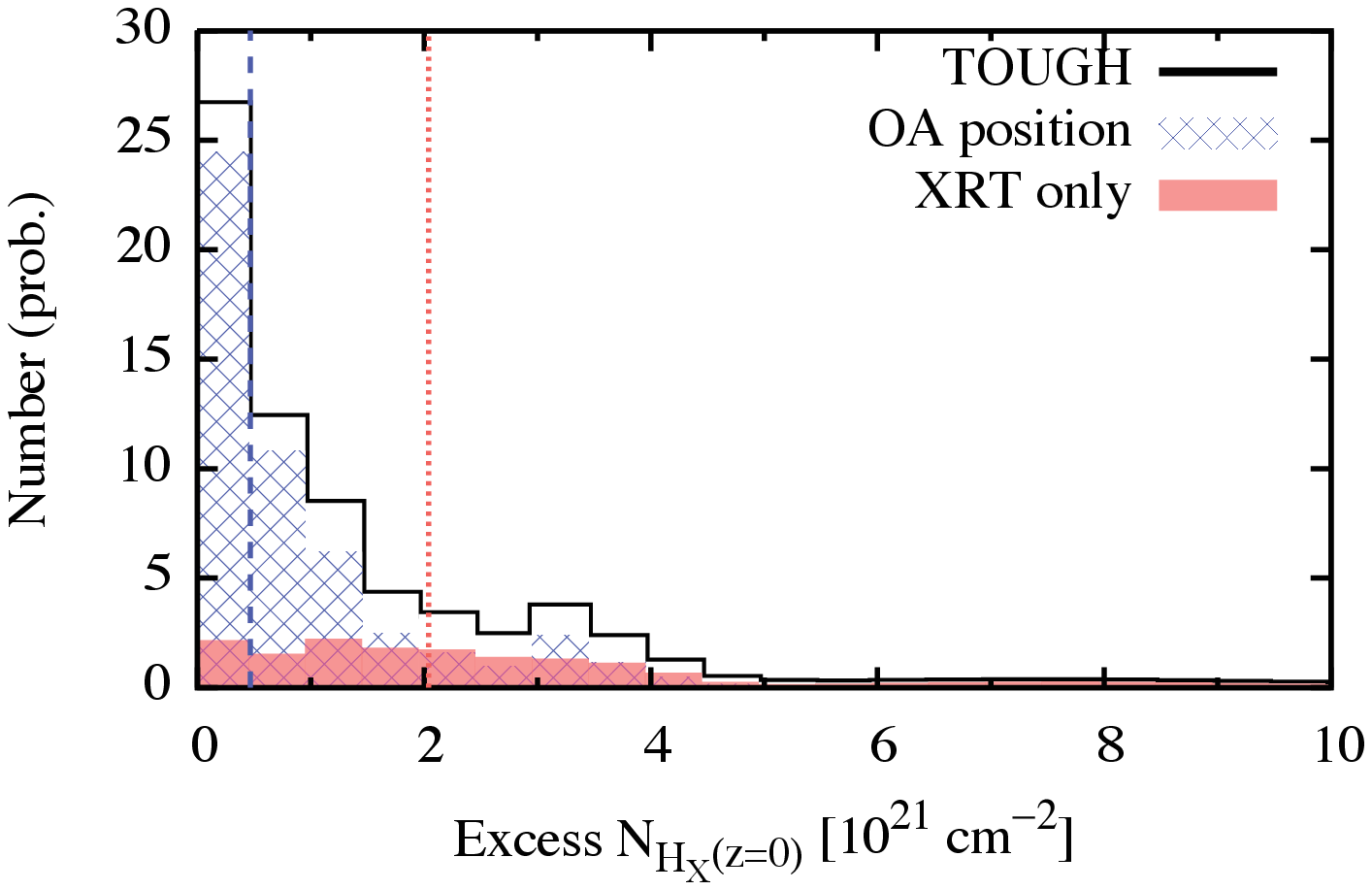}
\caption{Distributions of host galaxy $R$- (top) and $K_{s}$-band
(middle) magnitudes as well as X-ray absorptions in excess of
Galactic, determined at $z = 0$ (bottom) for the TOUGH sample
(open histogram). Subsets of galaxies selected with a localization
from an optical/NIR afterglow (OA) and a localization from \swift/XRT only are
shown as hatched and solid histograms, respectively. The corresponding
median values (or upper limit to the median as indicated by arrows) are shown
as dashed or dotted lines. The histograms were created from the sum
of the normalized probability distributions for each object derived
from their uncertainties or upper limits. For the galaxy magnitudes the upper
limit probabilities were evenly distributed in magnitude space from the upper
limit to a lower cutoff value: $R=30$ and $K_{s}=26$. While the height of
the histogram at low values depends on this choice, the histogram area does 
not.
\label{f:R_hist}}
\end{figure}

Figure~\ref{f:R_hist} highlights the difference between systems with and 
without an optical/NIR afterglow. There is a significant dearth of XRT-only 
galaxies at very faint magnitudes in the $R$ band. This is also reflected in 
the median $R$-band magnitude of $25.8\pm0.3$ for galaxies hosting an optically/NIR-detected
afterglow, as opposed to $24.4\pm0.4$ for those with undetected ones (Table 4). XRT-only 
localised host galaxies are also relatively brighter in $K_{s}$ (median magnitude 
of $K_{s}=21.1\pm 0.3$ opposed to $K_{s}>22.5$ for galaxies hosting GRBs with optical/NIR 
afterglows). A clue to the origin of these remarkable differences comes from 
the lower panel of Figure~\ref{f:R_hist} which demonstrates that XRT-only 
hosts have equivalent X-ray absorption column densities at $z=0$ extending to 
much higher values \citep[see also][]{2012arXiv1205.4036K}. These differences clearly 
demonstrate that statistical studies based on optically selected GRBs will be 
strongly biased. 

We note that our procedure of identifying the host as the brightest galaxy 
inside the XRT error circle naturally will lead to a bias toward brighter 
galaxies if any of the identifications are incorrect. To quantify this bias we 
tested our host identification procedure in the XRT error circles of GRBs
with an optical/NIR afterglow. We found that in 6 out of 50 cases ($12\%$) 
we would misidentify the host (this would indicate that about 
1--2 of the galaxies in Figure~\ref{f:malesani1} may have been misidentified). 
This procedure led to a biased median $R$-band magnitude of $25.5\pm0.2$ for 
the XRT error circles of GRB with optical/NIR afterglows but XRT-only 
identified 
host galaxies. Hence, the difference between host galaxies with and without 
optical/NIR afterglows remains significant.

Regarding systems with both $R$ and $K_{s}$ detections, we find that the median 
magnitude of hosts of GRBs with or without optical/NIR afterglows is consistently 
$R\approx24.3\pm0.3$. However, the median colors of XRT-only GRB host galaxies 
detected in the $K_{s}$ band are significantly redder, $R-K_{s}=3.7\pm0.3$, than 
those for which an optical/NIR afterglow was detected, $R-K_{s}=2.8\pm0.1$. While this 
difference in color could partly be a redshift effect, their larger X-ray 
absorption column densities suggest that GRBs without an optical/NIR afterglow 
suffer stronger attenuation due to dust and that dust extinction causes optical/NIR GRB 
afterglows to drop out of detection. This is in line with other studies which
find that host galaxies of dark GRBs are redder and brighter (more massive) 
than those of GRBs with optical afterglows 
\citep{2011A&A...534A.108K,2012MNRAS.421...25S}. A more detailed discussion of 
the $R-K_{s}$ distribution as a function of redshift and a comparison to other 
high-redshift galaxies are presented in D. Malesani (2012, in preparation).

\subsection{Star formation in GRB host galaxies\label{sf}}

Motivated by the fact that, before our survey, five out of a possible five 
hosts in a pre-\swift{} GRB sample were found to be LAEs
\citep{2002A&A...388..425F,2003A&A...406L..63F,2005MNRAS.362..245J},
we have also performed spectroscopic observations of systems with known 
redshift in the redshift range 1.8--4.5  where the Ly$\alpha$ emission line 
can be best
observed.  Remarkably, as reported by \citet{2012arXiv1205.3779M}, only 7 out of 20 galaxies 
targeted have secure detection of Ly$\alpha$ emission while upper limits 
on the Ly$\alpha$ equivalent width are derived for another 7 galaxies from 
detecting the continuum in the spectra 
(Figure~\ref{f:malesani2}). The origin of the lower (but more representative) 
success rate is likely a combination of lower intrinsic luminosities of the 
galaxies (and hence lower average Ly$\alpha$ luminosities) and lower 
equivalent widths, at least partly due to higher intrinsic extinction in the 
host galaxy. The Ly$\alpha$ emission
is still stronger than in flux-limited samples 
of LBGs.
Moreover, by comparison to the afterglow redshifts, we find that
the velocity centroid of the Ly$\alpha$ line is redshifted by 
200--600 km s$^{-1}$ with respect to the systemic velocity, indicative of 
outflows in the host galaxies 
\citep[e.g.,][]{2003ApJ...584...45A,2006A&A...460..397V,2009ApJ...704.1640L}. 

In order to assess the SFRs of GRB hosts, we have 
performed radio observations of TOUGH GRB hosts at $z<1$ \citep{2012arXiv1205.4239M}. 
This redshift limit was chosen to obtain meaningful limits on SFRs. We did 
not detect any TOUGH GRB hosts, which indicates that their average SFR is 
below $\sim15\,M_\odot~\mbox{yr}^{-1}$. We also found that at least $65$\% of 
GRB hosts at $z<1$ have $\mbox{SFR}<100\,M_\odot\mbox{ yr}^{-1}$ and that at 
least $92$\% of them have $A_V<3$ mag. The obtained limits allowed us to 
conclude that the distribution of SFRs and dust attenuation of GRB hosts at 
$z<1$ is consistent with that of other star-forming galaxies at similar 
redshifts.

Finally, in S. Schulze et al.\ (2012, in preparation)
the magnitudes are used to determine the luminosity 
function of GRB-selected galaxies. In particular, determining the slope of the 
faint end power-law part of the luminosity function is important in 
determining whether such galaxies constitute a major fraction of the star 
formation in the universe 
\citep{2001A&A...373..796F,2002A&A...388..425F,2005MNRAS.362..245J,2012ApJ...752...62J,2012ApJ...744...95R,2012ApJ...754...46T}.

\section{Summary\label{conclusion}}

Long-duration GRBs provide a complementary view of the nature of  ``typical"
star-forming galaxies over most of cosmic history. However,  to date, most
samples of GRB hosts have been heterogeneous and had a strong bias toward GRBs
with detected optical afterglows. In this paper we have  described a large,
homogeneous sample of hosts targeted with the VLT, using X-ray positions where 
no optical/NIR afterglow was found. We report the main catalog
in 
Table~3.
We detect the majority of the hosts, namely 80\%, in 
the $R$-band down to a detection limit of $\sim27.5$, including 17 systems at 
$z>2$. Within the TOUGH sample, we detect 54 hosts, of which 32 were not 
previously detected. GRB host galaxies cover a wide range of luminosities 
from  $<0.1 L^*$ to $L^*$. From the $K_{s}$-band imaging we confirm the result 
of more biased studies that GRB host galaxies tend to be blue. We also find 
that galaxies with no optical/NIR afterglow are significantly brighter and redder 
than those with an optical/NIR afterglow, highlighting the importance of TOUGH in 
being X-ray selected. With our newly reported redshifts, the redshift 
distribution consisting of 77\% of the GRBs, now appears smooth with a median 
redshift above 2.

Our sample will form the basis for statistical studies of GRB host galaxies, 
such as their luminosity function, colors and SFRs, redshift 
distribution, Ly$\alpha$ properties, radio emission, and X-ray absorption 
properties. 

Future observational work will focus on completing the sample, in particular
in redshift, and to complement it at other wavelengths, in particular, radio, 
sub-mm (ALMA, {\it Herschel}), and NIR/MIR ({\it Spitzer, HST, JWST}). It is 
also 
useful for selecting subsamples for further study, such as SED studies or 
emission-line diagnostics, e.g., metallicity determination. Finally, it is 
straightforward to enlarge in a well-defined way, in time or declination, as 
we are doing with VLT/X-shooter and Gemini.

\acknowledgements

We thank the members of the former GRACE collaboration as well as participants 
of the former EU FP5 Research Training Network ``Gamma-Ray Bursts: An Enigma 
and a Tool'', 
for their initial support of and continued interest in this project. We are 
particularly grateful to Paul M. Vreeswijk for his enthusiastic contributions 
to the project. We are grateful to Scott Barthelmy, Nat Butler, Phil Evans, 
Samantha Oates, and Patricia Schady for their assistance in defining our 
sample selection criteria. We acknowledge the use of Jochen Greiner's GRB
Web site. This work made use of data supplied by the UK Swift Science Data 
Centre at the University of Leicester. D.M. acknowledges support from the 
Instrument Center for Danish Astrophysics. P.J. and S.S. acknowledge support by 
a Project Grant from the Icelandic Research Fund. J.P.U.F. and B.M.-J. 
acknowledge support from the ERC-StG grant EGGS-278202. The research 
activities of J.G. are supported by the Spanish research programs 
AYA2008-03467/ESP and AYA2009-14000-C03-01. M.J.M. acknowledges the support of 
the Science and Technology Facilities Council. The Dark Cosmology Centre is 
funded by the Danish National Research Foundation.

Facilities: \facility{VLT:Antu(FORS2,ISAAC), VLT:Kueyen (FORS1,XSHOOTER), Swift(BAT,XRT)}

\clearpage

\bibliography{ms}

\clearpage

\input{table1}

\input{table2_new}

\input{table3_new}

\input{table4}

\end{document}

%% file: table1.tex
\begin{deluxetable}{lll}
\label{tab:criteria}
\tabletypesize{\scriptsize}
\tablecolumns{13}
\tablewidth{0pc}
\tablecaption{Sample Selection Criteria}
\tablehead {
\colhead{Criterion No.}	    &
\colhead{Short Description}      &
\colhead{Type of Constraint}      \\
}
\startdata
1  & Onboard \swift/BAT trigger                        & GRB properties\\
2  & Long GRBs: duration $ T_{90} > 2$ s               & GRB properties \\
3  & X-ray afterglow position available within 12 hr   & GRB properties, visibility, and slight biasing \\
4  & Galactic $A_V \leqslant 0.5$~mag                  & Visibility \\
5  & Sun distance $> 55$ deg                           & Visibility \\
6  & Not near bright stars                             & Visibility \\
7  & 2005 Mar 1 -- 2007 Aug 10                         & VLT sample\\ 
8  & $-70\degr < \delta < +27\degr$ (J2000.0)          & VLT sample\\ 
9  & Localization accuracy better than 2\arcsec        & Slight biasing
\enddata
\end{deluxetable}

%% file: table2_new.tex
\begin{deluxetable}{llllcllllrr}
\label{tab:sample}
\tabletypesize{\scriptsize}
\tablecolumns{13}
\tablewidth{0pc}
\tablecaption{The TOUGH Host Galaxy Sample}
\tablehead {
\colhead{GRB}        &
\colhead{Trigger No.} &
\colhead{R.A.}       &
\colhead{Decl.}      &
\colhead{Error}      &
\colhead{Band}       &
\colhead{Reference}       &
\colhead{$A_V$}      &
\colhead{$t_{\rm XRT}$}  &
\colhead{$T_{90}$}   &
\colhead{Peak Flux}  \\
\colhead{}           &
\colhead{}           &
\colhead{(J2000.0)}  &
\colhead{(J2000.0)}  &
\colhead{($''$)}     &
\colhead{}           &
\colhead{}           &
\colhead{(mag)}      &
\colhead{(s)}        &
\colhead{(s)}        &
\colhead{($\gamma$ cm$^{-2}$ s$^{-1}$)} \\
\colhead{(1)}        &
\colhead{(2)}        &
\colhead{(3)}        &
\colhead{(4)}        &
\colhead{(5)}        &
\colhead{(6)}        &
\colhead{(7)}        &
\colhead{(8)}        &
\colhead{(9)}        &
\colhead{(10)}        &
\colhead{(11)}       
}

\startdata
050315   & 111063  & 20:25:54.14  & $-$42:36:01.5  & 1.4  & $r$   & 1      & 0.16  & 84     &   95.6  &  1.89  \\
050318   & 111529  & 03:18:51.09  & $-$46:23:45.8  & 1.4  & $U$   & 2      & 0.05  & 3277   &   40.0  &  3.16  \\
050401   & 113120  & 16:31:28.80  & $+$02:11:14.5  & 3.5  & $V$   & 3      & 0.22  & 124    &   33.3  & 10.90  \\
050406   & 113872  & 02:17:52.09  & $-$50:11:17.1  & 1.9  & $U$   & 4      & 0.07  & 106    &    4.8  &  0.36  \\
050416A  & 114753  & 12:33:54.64  & $+$21:03:26.8  & 1.4  & UVW2  & 5      & 0.10  & 78     &    6.6  &  4.88  \\
050502B  & 116116  & 09:30:10.13  & $+$16:59:48.0  & 1.4  & $R$   & 6      & 0.10  & 63     &   16.6  &  1.40  \\
050525A  & 130088  & 18:32:32.67  & $+$26:20:21.6  & 1.5  & UVW2  & 7      & 0.32  & 125    &    8.8  & 41.00  \\
050714B  & 145994  & 11:18:47.71  & $-$15:32:49.0  & 1.5  & X     & 8      & 0.18  & 151    &   46.9  &  0.52  \\
050726   & 147788  & 13:20:11.96  & $-$32:03:51.6  & 1.4  & $V$   & 9      & 0.21  & 132    &   49.7  &  1.38  \\
050730   & 148225  & 14:08:17.18  & $-$03:46:18.1  & 1.4  & $B$   & 10,11  & 0.17  & 133    &  145.1  &  0.56  \\
050801   & 148522  & 13:36:35.32  & $-$21:55:42.7  & 1.5  & UVW2  & 12     & 0.32  & 61     &   19.4  &  1.46  \\
050803   & 148833  & 23:22:37.85  & $+$05:47:08.5  & 1.4  & X     & 13     & 0.25  & 152    &   88.1  &  0.95  \\
050819   & 151131  & 23:55:01.62  & $+$24:51:39.0  & 1.5  & X     & 14     & 0.41  & 141    &   37.7  &  0.40  \\
050820A  & 151207  & 22:29:38.12  & $+$19:33:37.0  & 1.4  & UVW1  & 15     & 0.15  & 80     &   26.0  &  2.45  \\
050822   & 151486  & 03:24:27.18  & $-$46:02:00.0  & 1.4  & X     & 16     & 0.05  & 96     &  104.1  &  2.23  \\
050824   & 151905  & 00:48:56.21  & $+$22:36:33.1  & 1.5  & UVW2  & 17     & 0.12  & 6089   &   24.8  &  0.49  \\
050904   & 153514  & 00:54:50.84  & $+$14:05:09.3  & 3.5  & $I$   & 18     & 0.20  & 161    &  181.7  &  0.64  \\
050908   & 154112  & 01:21:50.70  & $-$12:57:18.0  & 1.5  & $V$   & 19     & 0.08  & 106    &   18.3  &  0.70  \\
050915A  & 155242  & 05:26:44.81  & $-$28:00:59.4  & 1.4  & $H$   & 20     & 0.09  & 87     &   53.4  &  0.76  \\
050922B  & 156434  & 00:23:13.37  & $-$05:36:16.7  & 1.7  & X     & 21     & 0.12  & 342    &  156.3  &  1.10  \\
050922C  & 156467  & 21:09:33.08  & $-$08:45:30.2  & 1.4  & UVM2  & 15     & 0.34  & 108    &    4.5  &  7.04  \\
051001   & 157870  & 23:23:48.72  & $-$31:31:23.4  & 1.7  & X     & 22     & 0.05  & 192    &  190.6  &  0.48  \\
051006   & 158593  & 07:23:14.14  & $+$09:30:20.0  & 1.5  & X     & 23     & 0.22  & 113    &   26.0  &  1.70  \\
051016B  & 159994  & 08:48:27.85  & $+$13:39:20.4  & 1.4  & UVW2  & 24     & 0.12  & 75     &    4.0  &  1.28  \\
051117B  & 164279  & 05:40:43.38  & $-$19:16:27.2  & 1.7  & X     & 25     & 0.18  & 135    &    9.0  &  0.50  \\
060115   & 177408  & 03:36:08.32  & $+$17:20:43.1  & 1.4  & $R$   & 26     & 0.44  & 113    &  122.2  &  0.86  \\
060218   & 191157  & 03:21:39.69  & $+$16:52:01.6  & 1.4  & UVW2  & 27     & 0.47  & 153    & 2100.0  &  0.25  \\
060306   & 200638  & 02:44:22.88  & $-$02:08:54.7  & 1.4  & X     & 28     & 0.12  & 88     &   60.9  &  5.92  \\
060522   & 211117  & 21:31:44.89  & $+$02:53:10.1  & 1.4  & $R$   & 29     & 0.18  & 142    &   69.1  &  0.54  \\
060526   & 211957  & 15:31:18.33  & $+$00:17:05.4  & 1.4  & $B$   & 30,31  & 0.22  & 73     &  275.2  &  1.67  \\
060604   & 213486  & 22:28:55.04  & $-$10:54:56.1  & 1.4  & UVW1  & 32     & 0.14  & 109    &   96.0  &  0.34  \\
060605   & 213630  & 21:28:37.31  & $-$06:03:30.7  & 1.4  & $B$   & 33     & 0.16  & 93     &  539.1  &  0.47  \\
060607A  & 213823  & 21:58:50.41  & $-$22:29:47.0  & 1.4  & $U$   & 34     & 0.10  & 65     &  103.0  &  1.40  \\
060614   & 214805  & 21:23:32.12  & $-$53:01:36.6  & 1.4  & UVW2  & 35     & 0.07  & 91     &  109.2  & 11.50  \\
060707   & 217704  & 23:48:19.07  & $-$17:54:18.0  & 1.7  & $B$   & 36     & 0.07  & 122    &   66.7  &  1.08  \\
060708   & 217805  & 00:31:13.80  & $-$33:45:32.6  & 1.4  & UVW1  & 9      & 0.04  & 62     &   10.0  &  1.91  \\
060714   & 219101  & 15:11:26.46  & $-$06:33:58.8  & 1.4  & $B$   & 37     & 0.26  & 99     &  116.0  &  1.27  \\
060719   & 220020  & 01:13:43.70  & $-$48:22:50.6  & 1.4  & $K$   & 38     & 0.03  & 77     &   66.9  &  2.21  \\
060729   & 221755  & 06:21:31.79  & $-$62:22:12.5  & 1.4  & UVW2  & 39     & 0.18  & 124    &  113.0  &  1.14  \\
060805A  & 222683  & 14:43:43.47  & $+$12:35:11.2  & 1.6  & X     & 40     & 0.08  & 93     &    4.9  &  0.31  \\
060814   & 224552  & 14:45:21.36  & $+$20:35:09.2  & 1.4  & $K$   & 41     & 0.13  & 72     &  144.9  &  7.26  \\
060908   & 228581  & 02:07:18.42  & $+$00:20:32.2  & 1.4  & UVW1  & 9      & 0.10  & 72     &   18.8  &  3.09  \\
060912A  & 229185  & 00:21:08.13  & $+$20:58:18.5  & 1.4  & UWM2  & 9      & 0.17  & 109    &    5.0  &  8.27  \\
060919   & 230115  & 18:27:41.80  & $-$51:00:52.1  & 1.7  & X     & 42     & 0.24  & 87     &    9.0  &  2.31  \\
060923A  & 230662  & 16:58:28.14  & $+$12:21:37.9  & 1.5  & $K$   & 43     & 0.20  & 81     &   51.5  &  1.38  \\
060923B  & 230702  & 15:52:46.70  & $-$30:54:13.7  & 1.8  & X     & 44     & 0.49  & 114    &    8.9  &  1.43  \\
060923C  & 230711  & 23:04:28.28  & $+$03:55:28.1  & 3.5  & $J$   & 45     & 0.21  & 203    &   67.4  &  0.89  \\
060927   & 231362  & 21:58:12.05  & $+$05:21:49.0  & 1.6  & $I$   & 46     & 0.21  & 65     &   22.4  &  2.68  \\
061004   & 232339  & 06:31:10.81  & $-$45:54:24.2  & 1.4  & X     & 47     & 0.16  & 60     &    6.3  &  2.48  \\
061007   & 232683  & 03:05:19.60  & $-$50:30:02.4  & 1.4  & UVW2  & 48     & 0.07  & 80     &   75.7  & 14.70  \\
061021   & 234905  & 09:40:36.12  & $-$21:57:05.2  & 1.4  & UVW2  & 9      & 0.19  & 78     &   43.8  &  6.12  \\
061110A  & 238108  & 22:25:09.89  & $-$02:15:30.4  & 1.6  & $R$   & 49     & 0.30  & 69     &   44.5  &  0.48  \\
061110B  & 238174  & 21:35:40.46  & $+$06:52:32.9  & 1.7  & $R$   & 50     & 0.13  & 3195   &  132.8  &  0.47  \\
061121   & 239899  & 09:48:54.59  & $-$13:11:42.1  & 1.4  & UVW2  & 51     & 0.15  & 55     &   81.2  & 21.30  \\
070103   & 254532  & 23:30:13.80  & $+$26:52:34.4  & 1.5  & X     & 52     & 0.22  & 69     &   18.4  &  1.05  \\
070110   & 255445  & 00:03:39.38  & $-$52:58:28.1  & 1.4  & $U$   & 53     & 0.05  & 93     &   79.7  &  0.62  \\
070129   & 258408  & 02:28:00.98  & $+$11:41:03.4  & 1.4  & $R$   & 54     & 0.46  & 134    &  459.1  &  0.56  \\
070224   & 261880  & 11:56:06.57  & $-$13:19:48.8  & 1.6  & $R$   & 55     & 0.19  & 143    &   48.0  &  0.34  \\
070306   & 263361  & 09:52:23.29  & $+$10:28:55.5  & 1.4  & $H$   & 56     & 0.09  & 153    &  209.2  &  4.04  \\
070318   & 271019  & 03:13:56.77  & $-$42:56:47.3  & 1.4  & UVW2  & 57     & 0.06  & 64     &  131.5  &  1.64  \\
070328   & 272773  & 04:20:27.68  & $-$34:04:00.7  & 1.4  & X     & 58     & 0.12  & 88     &   72.1  &  4.22  \\
070330   & 273180  & 17:58:10.20  & $-$63:47:34.4  & 1.5  & $V$   & 59     & 0.21  & 68     &    6.6  &  0.88  \\
070419B  & 276212  & 21:02:49.77  & $-$31:15:48.7  & 1.5  & $V$   & 60     & 0.30  & 81     &  238.1  &  1.38  \\
070506   & 278693  & 23:08:52.31  & $+$10:43:20.8  & 1.8  & $B$   & 61     & 0.13  & 127    &    4.3  &  0.95  \\
070611   & 282003  & 00:07:58.12  & $-$29:45:20.4  & 1.8  & UVW1  & 62     & 0.04  & 3300   &   13.2  &  0.82  \\
070621   & 282808  & 21:35:10.08  & $-$24:49:03.1  & 1.4  & X     & 63     & 0.16  & 111    &   33.3  &  2.50  \\
070721B  & 285654  & 02:12:32.93  & $-$02:11:39.7  & 1.5  & $V$   & 64     & 0.10  & 92     &  336.9  &  1.57  \\
070802   & 286809  & 02:27:35.78  & $-$55:31:39.5  & 1.5  & $B$   & 65     & 0.09  & 138    &   15.8  &  0.40  \\
070808   & 287260  & 00:27:03.36  & $+$01:10:34.4  & 1.5  & X     & 66     & 0.08  & 114    &   58.4  &  1.94  \\
\hline
050412   & 114485  & 12:04:25.28  & $-$01:12:00.6  & 3.7  & X     & 67     & 0.07  & 99     &   26.5  &  0.49  \\
050603   & 131560  & 02:39:56.89  & $-$25:10:54.9  & 1.5  & $V$   & 68     & 0.09  & 39024  &   22.0  & 15.30  \\
060117   & 177666  & 21:51:36.23  & $-$59:58:39.3  & 1.5  & $R$   & 69     & 0.12  & \ldots &   16.9  & 48.80  \\
060223A  & 192059  & 03:40:49.56  & $-$17:07:49.7  & 1.5  & $V$   & 9      & 0.39  & 86     &   11.3  &  1.35  \\
060313   & 201487  & 04:26:28.46  & $-$10:50:41.2  & 1.4  & UVW2  & 70     & 0.23  & 79     &    0.7  & 10.80  \\
060505   & 208654  & 22:07:03.32  & $-$27:48:53.0  & 2.0  & $B$   & 71     & 0.07  & 51744  &    4.0  &  2.65  \\
060602A  & 213180  & 09:58:16.28  & $+$00:18:14.4  & 5.2  & $R$   & 72     & 0.08  & 155520 &   74.7  &  0.56  \\
070721A  & 285653  & 00:12:39.13  & $-$28:33:00.9  & 1.6  & X     & 73     & 0.05  & 86     &    3.4  &  0.68  \\
070810A  & 287364  & 12:39:51.24  & $+$10:45:03.2  & 1.4  & $V$   & 74     & 0.07  & 88     &    9.0  &  1.89  \\
080207   & 302728  & 13:50:02.93  & $+$07:30:07.9  & 1.4  & X     & 75     & 0.08  & 124    &  340.0  &  1.00  \\
\enddata

\tablecomments{Column 1: GRB name; Column 2: \textit{Swift} trigger number;
Columns 3--5:
X-ray position and error \citep{2007A&A...476.1401G,2009MNRAS.397.1177E};  the
position of GRB\,060117 is taken from \citet{2006A&A...454L.119J}; Column 6:
bluest
optical/NIR afterglow detection, excluding white and unfiltered measurements;
``X'' indicates no optical/NIR detection; Column 7: reference for the afterglow
detection; Column 8: Galactic $V$-band absorption \citep{1998ApJ...500..525S};
Column 9:
time to start of the XRT observation; Column 10: observed gamma-ray duration
$T_{90}$ \citep{2011ApJS..195....2S}; Column 11: peak photon flux over 1~s
timescale (15--150 keV; \citealt{2011ApJS..195....2S}). The horizontal line
separates the objects in the TOUGH sample from 10 extra systems that were
observed within our program.\\
References.
(1) \citet{2005ApJ...634..501B}; 
(2) \citet{2005ApJ...635.1187S}; 
(3) \citet{2009MNRAS.394..214K}; 
(4) \citet{2006ApJ...643..276S}; 
(5) \citet{2007AJ....133..122H}; 
(6) \citet{2011A&A...526A.154A}; 
(7) \citet{2006ApJ...637..901B}; 
(8) \citet{2005GCN..3613....1L}; 
(9) \citet{2009MNRAS.395..490O}; 
(10) \citet{2006A&A...460..415P}; 
(11) \citet{2007A&A...471...83P}; 
(12) \citet{2007MNRAS.377.1638D}; 
(13) \citet{2005GCN..3748....1B}; 
(14) \citet{2005GCN..3827....1K}; 
(15) \citet{2010MNRAS.401.2773S}; 
(16) \citet{2005GCN..3849....1B}; 
(17) \citet{2007MNRAS.377..273S}; 
(18) \citet{2005A&A...443L...1T}; 
(19) \citet{2005GCN..3960....1D}; 
(20) \citet{2005GCN..3984....1B}; 
(21) \citet{2005GCN..4008....1N}; 
(22) \citet{2005GCN..4051....1M}; 
(23) \citet{2005GCN..4062....1M}; 
(24) \citet{2005GCN..4107....1B}; 
(25) \citet{2005GCN..4282....1B}; 
(26) \citet{2006GCN..4510....1Y}; 
(27) \citet{2006Natur.442.1008C}; 
(28) \citet{2006GCN..4848....1A}; 
(29) \citet{2006GCN..5151....1D}; 
(30) \citet{2007ApJ...658..509D}; 
(31) \citet{2010A&A...523A..70T}; 
(32) \citet{2006GCN..5219....1B}; 
(33) \citet{2009A&A...497..729F}; 
(34) \citet{2008MNRAS.385..453Z}; 
(35) \citet{2007A&A...470..105M}; 
(36) \citet{2006GCN..5294....1S}; 
(37) \citet{2006GCN..5434....1A}; 
(38) \citet{2006GCN..5350....1M}; 
(39) \citet{2007ApJ...662..443G}; 
(40) \citet{2006GCN..5398....1Z}; 
(41) \citet{2006GCN..5455....1L}; 
(42) \citet{2006GCN..5575....1G}; 
(43) \citet{2008MNRAS.388.1743T}; 
(44) \citet{2006GCN..5590....1S}; 
(45) \citet{2006GCN..5604....1C}; 
(46) \citet{2007ApJ...669....1R}; 
(47) \citet{2006GCN..5691....1Z}; 
(48) \citet{2007MNRAS.380.1041S}; 
(49) \citet{2006GCN..5797....1C}; 
(50) \citet{2006GCN..5804....1M}; 
(51) \citet{2007ApJ...663.1125P}; 
(52) \citet{2007GCN..5988....1S}; 
(53) \citet{2007ApJ...665..599T}; 
(54) \citet{2007GCN..6055....1M}; 
(55) \citet{2007GCN..6142....1T}; 
(56) \citet{2008ApJ...681..453J}; 
(57) \citet{2008AIPC.1000..421C}; 
(58) \citet{2007GCN..6224....1M}; 
(59) \citet{2007GCN..6238....1K}; 
(60) \citet{2007GCN..6321....1D}; 
(61) \citet{2007GCN..6378....1L}; 
(62) \citet{2007GCN..6504....1L}; 
(63) \citet{2007GCN..6560....1S}; 
(64) \citet{2007GCN..6650....1D}; 
(65) \citet{2009ApJ...697.1725E}; 
(66) \citet{2007GCN..6718....1C}; 
(67) \citet{2007A&A...469..663M}; 
(68) \citet{2006ApJ...645..464G}; 
(69) \citet{2006A&A...454L.119J}; 
(70) \citet{2006ApJ...651..985R}; 
(71) \citet{2009ApJ...696..971X}; 
(72) \citet{2006GCN..5203....1J}; 
(73) \citet{2007GCN..6639....1Z}; 
(74) \citet{2007GCN..6735....1C}; 
(75) \citet{2008GCN..7264....1R}. 
}

\end{deluxetable}

%% file: table3_new.tex

\begin{deluxetable}{llllllll}
\label{tab:results}
\tabletypesize{\scriptsize}
\tablecolumns{13}
\tablewidth{0pc}
\tablecaption{Catalog of Survey Results}
\tablehead {
\colhead{GRB}	    &
\colhead{R.A.}      &
\colhead{Decl.}     &
\colhead{$R$}       &
\colhead{$K_{\rm s}$} &
\colhead{Redshift}  &
\colhead{Notes}     &
\colhead{Reference}      \\
\colhead{}	    &
\colhead{(J2000.0)} &
\colhead{(J2000.0)} &
\colhead{(mag)}	    &
\colhead{(mag)}	    &
\colhead{}	    &
\colhead{}	    &
\colhead{}	    \\
\colhead{(1)}	    &
\colhead{(2)}	    &
\colhead{(3)}	    &
\colhead{(4)}	    &
\colhead{(5)}	    &
\colhead{(6)}	    &
\colhead{(7)}	    &
\colhead{(8)}	    
}

\startdata
050315   & 20:25:54.189  & $-$42:36:02.16  & 24.43$\pm$0.15  & 21.89$\pm$0.36  & 1.9500       & A        & 1         \\
050318   & \nodata       & \nodata         & $>26.79$        & $>22.31$        & 1.4436       & A        & 1         \\
050401   & 16:31:28.794  & $+$02:11:14.33  & 26.15$\pm$0.31  & $>21.73$        & 2.8983       & A*       & 2,3       \\
050406   & 02:17:52.224  & $-$50:11:14.97  & 26.61$\pm$0.34  & $>21.25$        & 2.70         & P        & 4         \\
050416A  & 12:33:54.602  & $+$21:03:26.58  & 23.13$\pm$0.03  & 20.36$\pm$0.20  & 0.6528       & E        & 1,5       \\
050502B  & \nodata       & \nodata         & $>26.79$        & $>21.81$        & 5.2          & P        & 6         \\
050525A  & 18:32:32.590  & $+$26:20:22.60  & 25.95$\pm$0.34  & $>21.23$        & 0.606        & AE       & 7         \\
050714B  & 11:18:47.717  & $-$15:32:49.01  & 25.45$\pm$0.20  & 20.39$\pm$0.17  & \nodata      & \nodata  & \nodata   \\
050726   & \nodata       & \nodata         & $>26.14$        & $>21.28$        & \nodata      & \nodata  & \nodata   \\
050730   & \nodata       & \nodata         & $>27.20$        & $>21.36$        & 3.96855      & A*       & 8,9,3,10  \\
050801   & \nodata       & \nodata         & $>26.78$        & $>21.70$        & 1.38         & P        & 11,12     \\
050803   & 23:22:37.850  & $+$05:47:08.90  & 26.18$\pm$0.33  & $>21.50$        & \nodata      & \nodata  & \nodata   \\
050819   & 23:55:01.598  & $+$24:51:38.16  & 24.10$\pm$0.09  & \nodata         & \bf 2.5043   & E        & 13        \\
050820A  & 22:29:38.114  & $+$19:33:36.61  & 24.83$\pm$0.18  & $>21.18$        & 2.61469      & A*       & 14,15     \\
050822   & 03:24:27.214  & $-$46:01:59.43  & 24.19$\pm$0.08  & 20.78$\pm$0.20  & \bf 1.434    & E        & 16        \\
050824   & 00:48:56.151  & $+$22:36:32.80  & 24.23$\pm$0.16  & $>21.41$        & 0.8278       & AE       & 17,3      \\
050904   & \nodata       & \nodata         & \nodata         & \nodata         & 6.295        & A        & 18        \\
050908   & 01:21:50.730  & $-$12:57:17.30  & 27.76$\pm$0.45  & $>21.91$        & 3.3467       & A*       & 3,19      \\
050915A  & 05:26:44.843  & $-$28:00:59.62  & 24.56$\pm$0.16  & 20.69$\pm$0.24  & \bf 2.5273   & E        & 13        \\
050922B  & 00:23:13.345  & $-$05:36:18.31  & 26.18$\pm$0.30  & $>22.17$        & \nodata      & \nodata  & \nodata   \\
050922C  & \nodata       & \nodata         & $>26.34$        & $>21.47$        & 2.1992       & A*       & 20,21     \\
051001   & 23:23:48.800  & $-$31:31:23.45  & 24.36$\pm$0.13  & 21.13$\pm$0.35  & \bf 2.4296   & E        & 13        \\
051006   & 07:23:14.109  & $+$09:30:19.96  & 22.99$\pm$0.07  & \nodata         & \bf 1.059    & E        & 16        \\
051016B  & 08:48:27.844  & $+$13:39:20.07  & 22.64$\pm$0.03  & 20.04$\pm$0.20  & 0.9364       & E        & 22        \\
051117B  & 05:40:43.287  & $-$19:16:26.31  & 21.06$\pm$0.02  & 17.75$\pm$0.05  & \bf 0.481    & E        & 16        \\
060115   & 03:36:08.355  & $+$17:20:42.65  & 27.14$\pm$0.53  & $>21.27$        & 3.5328       & A*       & 3         \\
060218   & 03:21:39.683  & $+$16:52:02.09  & 19.98$\pm$0.01  & 17.94$\pm$0.09  & 0.03351      & E        & 23,24,25,26,27 \\
060306   & 02:44:22.930  & $-$02:08:54.55  & 24.09$\pm$0.08  & 19.95$\pm$0.17  & \nodata      & \nodata  & \nodata   \\
060522   & \nodata       & \nodata         & $>26.64$        & $>23.15$        & 5.110        & A        & 28        \\
060526   & \nodata       & \nodata         & $>27.02$        & $>22.11$        & 3.2213       & A        & 29,20,30  \\
060604   & 22:28:55.029  & $-$10:54:55.72  & 25.52$\pm$0.18  & $>21.37$        & \bf 2.1357   & AE       & 3,13      \\
060605   & \nodata       & \nodata         & $>26.54$        & $>21.12$        & 3.773        & A        & 31        \\
060607A  & \nodata       & \nodata         & $>27.92$        & $>21.86$        & 3.0749       & A        & 15        \\
060614   & 21:23:32.103  & $-$53:01:36.20  & 22.53$\pm$0.06  & 20.36$\pm$0.20  & 0.125        & E        & 32,33     \\
060707   & 23:48:19.061  & $-$17:54:17.72  & 24.86$\pm$0.06  & $>19.71$        & 3.4240       & A        & 20,3      \\
060708   & 00:31:13.748  & $-$33:45:32.73  & 26.76$\pm$0.28  & $>21.04$        & 1.92         & P        & 11        \\
060714   & 15:11:26.436  & $-$06:33:58.24  & 26.45$\pm$0.28  & $>21.40$        & 2.7108       & AE*      & 20,3      \\
060719   & 01:13:43.706  & $-$48:22:51.31  & 24.62$\pm$0.12  & 21.25$\pm$0.27  & \bf 1.5320   & E        & 13        \\
060729   & 06:21:31.794  & $-$62:22:12.53  & 23.71$\pm$0.04  & 21.20$\pm$0.31  & 0.5428       & A        & 3         \\
060805A  & 14:43:43.381  & $+$12:35:10.28  & 23.48$\pm$0.03  & 20.69$\pm$0.15  & \nodata      & \nodata  & \nodata   \\
060805A  & 14:43:43.488  & $+$12:35:12.56  & 25.11$\pm$0.14  & 21.66$\pm$0.38  & \nodata      & \nodata  & \nodata   \\
060814   & 14:45:21.347  & $+$20:35:10.72  & 22.85$\pm$0.11  & 20.06$\pm$0.10  & \bf 1.9229   & E        & 16,13,34  \\
060908   & 02:07:18.406  & $+$00:20:31.55  & 25.53$\pm$0.18  & 22.53$\pm$0.43  & \bf 1.8836   & AE       & 35,3      \\
060912A  & 00:21:08.127  & $+$20:58:17.66  & 22.68$\pm$0.04  & 19.86$\pm$0.09  & 0.937        & E        & 36        \\
060919   & 18:27:41.790  & $-$51:00:50.94  & 25.79$\pm$0.26  & 21.41$\pm$0.38  & \nodata      & \nodata  & \nodata   \\
060923A  & 16:58:28.175  & $+$12:21:38.96  & 26.07$\pm$0.24  & $>21.25$        & \nodata      & \nodata  & \nodata   \\
060923B  & 15:52:46.566  & $-$30:54:14.57  & 24.34$\pm$0.16  & 20.39$\pm$0.19  & \nodata      & \nodata  & \nodata   \\
060923B  & 15:52:46.583  & $-$30:54:12.70  & 24.40$\pm$0.16  & 20.59$\pm$0.20  & \nodata      & \nodata  & \nodata   \\
060923C  & 23:04:28.266  & $+$03:55:29.29  & 25.45$\pm$0.18  & 21.73$\pm$0.34  & \nodata      & \nodata  & \nodata   \\
060927   & \nodata       & \nodata         & \nodata         & $>21.76$        & 5.4636       & A        & 37,3      \\
061004   & \nodata       & \nodata         & $>25.76$        & $>21.24$        & \nodata      & \nodata  & \nodata   \\
061007   & 03:05:19.587  & $-$50:30:02.43  & 24.40$\pm$0.17  & 21.06$\pm$0.37  & 1.2622       & AE       & 38,3      \\
061021   & 09:40:36.124  & $-$21:57:05.05  & 24.44$\pm$0.06  & $>21.14$        & \bf 0.3463   & AE       & 16,3      \\
061110A  & 22:25:09.843  & $-$02:15:31.12  & 25.65$\pm$0.26  & $>21.41$        & 0.7578       & E        & 3         \\
061110B  & 21:35:40.400  & $+$06:52:33.91  & 26.04$\pm$0.29  & $>21.24$        & 3.4344       & A*       & 3         \\
061121   & 09:48:54.564  & $-$13:11:43.09  & 22.75$\pm$0.03  & 20.14$\pm$0.19  & 1.3145       & A*       & 3         \\
070103   & 23:30:13.793  & $+$26:52:35.32  & 24.18$\pm$0.14  & 21.07$\pm$0.36  & \bf 2.6208   & E        & 13        \\
070110   & 00:03:39.268  & $-$52:58:27.12  & 25.02$\pm$0.11  & 22.41$\pm$0.50  & 2.3521       & AE*      & 3         \\
070129   & 02:28:00.915  & $+$11:41:04.56  & 24.39$\pm$0.12  & 21.01$\pm$0.18  & \bf 2.3384   & E        & 13        \\
070224   & 11:56:06.637  & $-$13:19:48.33  & 25.96$\pm$0.31  & $>21.20$        & \nodata      & \nodata  & \nodata   \\
070306   & 09:52:23.310  & $+$10:28:55.28  & 22.90$\pm$0.08  & 19.70$\pm$0.10  & \bf 1.49594  & E        & 39,16     \\
070318   & 03:13:56.800  & $-$42:56:46.25  & 24.43$\pm$0.11  & $>21.04$        & 0.8397       & A        & 3,40      \\
070328   & 04:20:27.615  & $-$34:04:00.63  & 24.43$\pm$0.13  & $>20.98$        & \nodata      & \nodata  & \nodata   \\
070330   & \nodata       & \nodata         & $>26.15$        & $>21.19$        & \nodata      & \nodata  & \nodata   \\
070419B  & 21:02:49.801  & $-$31:15:48.82  & 25.23$\pm$0.20  & $>21.22$        & \bf 1.9591   & E        & 13        \\
070506   & 23:08:52.392  & $+$10:43:21.00  & 26.10$\pm$0.22  & $>21.36$        & 2.3090       & A        & 3         \\
070611   & \nodata       & \nodata         & $>27.09$        & $>21.75$        & 2.0394       & A        & 3         \\
070621   & 21:35:10.061  & $-$24:49:02.18  & 25.77$\pm$0.23  & $>21.51$        & \nodata      & \nodata  & \nodata   \\
070721B  & 02:12:32.950  & $-$02:11:40.80  & 27.53$\pm$0.44  & $>22.38$        & 3.6298       & A*       & 3         \\
070802   & 02:27:35.722  & $-$55:31:38.76  & 25.11$\pm$0.21  & 21.27$\pm$0.41  & 2.4541       & A*       & 41,3      \\
070808   & 00:27:03.310  & $+$01:10:35.81  & 26.71$\pm$0.33  & 21.77$\pm$0.37  & \nodata      & \nodata  & \nodata   \\
\hline
050412   & 12:04:25.460  & $-$01:12:00.05  & 25.55$\pm$0.19  & $>21.40$        & \nodata      & \nodata  & \nodata   \\
050603   & \nodata       & \nodata         & $>26.61$        & $>22.66$        & \nodata      & \nodata  & \nodata   \\
060117   & \nodata       & \nodata         & $>26.28$        & \nodata         & \nodata      & \nodata  & \nodata   \\
060223A  & \nodata       & \nodata         & $>26.25$        & $>21.24$        & 4.406        & A        & 28        \\
060313   & \nodata       & \nodata         & \nodata         & $>21.13$        & \nodata      & \nodata  & \nodata   \\
060505   & 22:07:03.500  & $-$27:48:55.57  & 17.90$\pm$0.00  & 16.22$\pm$0.01  & 0.0889       & E        & 42,43,44  \\
060602A  & 09:58:16.750  & $+$00:18:12.56  & 23.88$\pm$0.08  & 20.56$\pm$0.27  & \bf 0.787    & E        & 45        \\
070721A  & 00:12:39.048  & $-$28:33:00.14  & 25.16$\pm$0.17  & \nodata         & \nodata      & \nodata  & \nodata   \\
070721A  & 00:12:39.139  & $-$28:33:00.98  & 23.14$\pm$0.15  & \nodata         & \nodata      & \nodata  & \nodata   \\
070810A  & \nodata       & \nodata         & $>26.73$        & $>21.93$        & 2.17         & A        & 46        \\
080207   & 13:50:02.997  & $+$07:30:07.44  & 25.84$\pm$0.18  & \nodata         & \bf 2.0858   & E        & 13        \\
\enddata

\tablecomments{Column1: GRB name; Columns 2 and 3: host coordinates;
Column 4: $R$-band Vega
magnitude; Column 5: $K_{\rm s}$-band Vega magnitude; Column 6: redshift; numbers in
boldface were determined from our program; Column 7: source of the redshift:
(A)bsorption features in the afterglow spectrum (*, with fine-structure
transitions), (E)mission lines from the host, or (P)hotometric redshift of the
afterglow; Column 8: redshift reference. The horizontal line separates the bursts in
the TOUGH sample from 10 extra hosts that were observed within our program.\\
References.
(1) \citet{2005ApJ...634..501B}; 
(2) \citet{2006ApJ...652.1011W}; 
(3) \citet{2009ApJS..185..526F}; 
(4) \citet{2006ApJ...643..276S}; 
(5) \citet{2007ApJ...661..982S}; 
(6) \citet{2011A&A...526A.154A}; 
(7) \citet{2006ApJ...642L.103D}; 
(8) \citet{2005ApJ...634L..25C}; 
(9) \citet{2005A&A...442L..21S}; 
(10) \citet{2007A&A...467..629D}; 
(11) \citet{2009MNRAS.395..490O}; 
(12) \citet{2007MNRAS.377.1638D}; 
(13) \citet{2012arXiv1205.4036K}; 
(14) \citet{2007ApJS..168..231P}; 
(15) \citet{2008A&A...491..189F}; 
(16) \citet{2012ApJ...752...62J}; 
(17) \citet{2007A&A...466..839S}; 
(18) \citet{2006Natur.440..184K}; 
(19) \citet{2005GCN..3971....1P}; 
(20) \citet{2006A&A...460L..13J}; 
(21) \citet{2008A&A...492..775P}; 
(22) \citet{2005GCN..4186....1S}; 
(23) \citet{2006ApJ...643L..99M}; 
(24) \citet{2006A&A...454..503S}; 
(25) \citet{2006ApJ...645L..21M}; 
(26) \citet{2006Natur.442.1011P}; 
(27) \citet{2007A&A...464..529W}; 
(28) \citet{2007ApJ...671..272C}; 
(29) \citet{2006GCN..5170....1B}; 
(30) \citet{2010A&A...523A..70T}; 
(31) \citet{2009A&A...497..729F}; 
(32) \citet{2006Natur.444.1053G}; 
(33) \citet{2006Natur.444.1050D}; 
(34) \citet{2012ApJ...749...68S}; 
(35) \citet{2012arXiv1205.3779M}; 
(36) \citet{2007MNRAS.378.1439L}; 
(37) \citet{2007ApJ...669....1R}; 
(38) \citet{2006GCN..5715....1O}; 
(39) \citet{2008ApJ...681..453J}; 
(40) \citet{2007GCN..6217....1C}; 
(41) \citet{2009ApJ...697.1725E}; 
(42) \citet{2007ApJ...662.1129O}; 
(43) \citet{2006Natur.444.1047F}; 
(44) \citet{2008ApJ...676.1151T}; 
(45) \citet{2007GCN..6997....1J}; 
(46) \citet{2007GCN..6741....1T}. 
}

\end{deluxetable}

%% file: table4.tex
\begin{deluxetable}{llllll}
\label{tab:median}
\tabletypesize{\scriptsize}
\tablecolumns{5}
\tablewidth{0pc}
\tablecaption{Median Properties of TOUGH and TOUGH Subsamples}
\tablehead {
\colhead{Sample} &
\colhead{$R$}    &
\colhead{$K_s$}    &
\colhead{$\langle{z}\rangle_{\rm low}$}   &
\colhead{Median $z$}   &
\colhead{$\langle{z}\rangle_{\rm high}$}  \\
\colhead{(1)}    &
\colhead{(2)}    &
\colhead{(3)}    &
\colhead{(4)}    &
\colhead{(5)}    &
\colhead{(6)}    \\
}
\startdata
TOUGH    &  $25.52\pm 0.23$  &  $>22.53$       &  1.50 &\nodata          & 2.53   \\
\hline
FORS     &  $23.59\pm 0.51$  &  $20.78\pm 1.30$ &\nodata& $1.25\pm 0.29$ & \nodata   \\
X-shooter&  $24.48\pm 0.15$  &  $21.13\pm 0.27$ &\nodata& $2.24\pm 0.15$ & \nodata  \\
OA       &  $25.80\pm 0.31$  &  $>22.53$        &  1.95 &\nodata         & 2.32   \\
XRT      &  $24.36\pm 0.40$  &  $21.13\pm 0.29$ &  \nodata       & \nodata & \nodata \\
$z$      &  $25.11\pm 0.31$  &  $>22.53$        &\nodata& $2.14\pm 0.18$ & \nodata   \\
No $z$   &  $25.88\pm 0.24$  &  $>21.77$        &  \nodata       & \nodata & \nodata \\
$R$      &  $24.73\pm 0.21$  &  $22.15\pm 0.69$ &  1.29 &\nodata         & 2.39   \\
No $R$   &   \nodata         &    \nodata       &  3.07 &\nodata         & 3.97   \\
$K_s$      &  $24.27\pm 0.28$  &  $20.74\pm 0.16$ &  1.10&\nodata          & 1.94   \\
No $K_s$   &  $26.61\pm 0.43$  &    \nodata       &  2.14 &\nodata         & 3.35   \\
\enddata
\tablecomments{Column 1: TOUGH sample or subsample; Column 2: 
median $R$ magnitude of
subsample; Column 3: median $K_s$ magnitude of subsample;
Column 4: minimum median redshift of subsample, assuming all unknown redshifts 
are less than the minimum redshift; Column 5:
median redshift; Column 6: maximum median redshift of subsample, assuming 
all unknown redshifts are larger than the maximum redshift.\\
Errors in the median are estimated as 
1.4826 $\times$ MAD/$\sqrt{N-1}$, where MAD 
is the median absolute deviation, $N$ is the number of systems used to 
determine the median and 1.4826 is a scale factor that applies to Gaussian 
distributions.\\
The subsamples are the following:
FORS: the six well-defined FORS emission-line redshifts, i.e., excluding the 
dropout redshifts (GRBs 050822, 051006, 051117B, 060908, 061021, and 070306);
X-shooter: the nine new X-shooter redshifts;
OA: systems with an optical/NIR afterglow;
XRT: systems with no optical/NIR afterglow;
$z$: systems with a known redshift;
No $z$: systems without a known redshift;
$R$: systems with a measured $R$ magnitude;
No $R$: systems without a measured $R$ magnitude (including the high-redshift
systems GRB 050904 and GRB 060927 which were not observed);
$K_s$: systems with a measured $K_s$ magnitude;
No $K_s$: systems without a measured $K_s$ magnitude (including the high-redshift GRB 050904 but excluding GRB 050819 and GRB 051006 which were not observed).}
\end{deluxetable}